\documentclass[letterpaper,twocolumn,10pt]{article}
\usepackage{zhanggroup}
\usepackage[absolute]{textpos}
\usepackage{multirow,colortbl}
\usepackage{epsfig,endnotes,xspace}
\usepackage{amsmath,amsthm,amsfonts}
\definecolor{mygray}{gray}{.9}
\usepackage{bbm}

\newcommand{\mypara}[1]{\smallskip\noindent\textbf{#1.}\xspace}
\newcommand{\tuple}[1]{\ensuremath{\langle #1 \rangle}}
\newcommand{\calN}{\mathcal{N}\xspace}

\newcommand{\scratch}{\ensuremath{\mathsf{Scratch}}\xspace}
\newcommand{\sisa}{\ensuremath{\mathsf{SISA}}\xspace}
\newcommand{\dd}{\ensuremath{\mathsf{DirectDiff}}\xspace}
\newcommand{\sd}{\ensuremath{\mathsf{SortedDiff}}\xspace}
\newcommand{\dc}{\ensuremath{\mathsf{DirectConcat}}\xspace}
\newcommand{\scc}{\ensuremath{\mathsf{SortedConcat}}\xspace}
\newcommand{\ed}{\ensuremath{\mathsf{EucDist}}\xspace}
\newcommand{\method}{our attack\xspace}
\newcommand{\mlleak}{classical membership inference\xspace}

\newcommand{\model}[2]{\ensuremath{\mathcal{M}_{#1}^{#2}}\xspace}
\newcommand{\dset}[2]{\ensuremath{\mathcal{D}_{#1}^{#2}}\xspace}
\newcommand{\post}[2]{\ensuremath{\mathbb{P}_{#1}^{#2}}\xspace}
\newcommand{\feat}[1]{\ensuremath{\mathcal{F}_{#1}}\xspace}

\newcommand{\so}{\ensuremath{S_o}\xspace}
\newcommand{\su}{\ensuremath{S_u}\xspace}
\newcommand{\sr}{\ensuremath{S_r}\xspace}
\newcommand{\too}{\ensuremath{T_o}\xspace}
\newcommand{\tu}{\ensuremath{T_u}\xspace}
\newcommand{\tr}{\ensuremath{T_r}\xspace}

\newcommand{\target}{target\xspace}
\newcommand{\unlearned}{unlearned\xspace}

\newcommand{\imprate}{DegRate\xspace}
\newcommand{\betrate}{DegCount\xspace}
\newcommand{\DegCount}{Degradation Count\xspace}
\newcommand{\DegRate}{Degradation Rate\xspace}

\begin{document}

\begin{textblock}{15}(2,1)
To Appear in 2021 ACM SIGSAC Conference on Computer and Communications Security, November 2021
\end{textblock}

\title{When Machine Unlearning Jeopardizes Privacy}
\date{}

\author{
Min Chen\textsuperscript{1}\thanks{Min and Zhikun contributed equally to the paper.}\ \ \
Zhikun Zhang\textsuperscript{1}\textsuperscript{$\ast$}\ \ \
Tianhao Wang\textsuperscript{$^{2,3}$}\thanks{Tianhao did most of the work while at Purdue University.}\ \ \
Michael Backes\textsuperscript{1}\ \ \ 
\\
Mathias Humbert\textsuperscript{4}\ \ \
Yang Zhang\textsuperscript{1}
\\
\textsuperscript{1}\textit{CISPA Helmholtz Center for Information Security} \ \ \ 
\textsuperscript{2}\textit{Carnegie Mellon University} \ \ \
\\
\textsuperscript{3}\textit{University of Virginia} \ \ \
\textsuperscript{4}\textit{University of Lausanne} \ \ \ 
}

\maketitle

\begin{abstract}
The right to be forgotten states that a data owner has the right to erase their data from an entity storing it. In the context of machine learning (ML), the right to be forgotten requires an ML model owner to remove the data owner's data from the training set used to build the ML model, a process known as \textit{machine unlearning}. 
While originally designed to protect the privacy of the data owner, we argue that machine unlearning may leave some imprint of the data in the ML model and thus create unintended privacy risks.
In this paper, we perform the first study on investigating the unintended information leakage caused by machine unlearning. We propose a novel membership inference attack that leverages the different outputs of an ML model's two versions to infer whether a target sample is part of the training set of the original model but out of the training set of the corresponding unlearned model. Our experiments demonstrate that the proposed membership inference attack achieves strong performance. More importantly, we show that our attack in multiple cases outperforms the classical membership inference attack on the original ML model, which indicates that machine unlearning can have counterproductive effects on privacy. We notice that the privacy degradation is especially significant for well-generalized ML models where classical membership inference does not perform well. We further investigate four mechanisms to mitigate the newly discovered privacy risks and show that releasing the predicted label only, temperature scaling, and differential privacy are effective. We believe that our results can help improve privacy protection in practical implementations of machine unlearning.\footnote{Our code is available at \url{https://github.com/MinChen00/UnlearningLeaks}.}
\end{abstract}

\section{Introduction}
\label{sec:intro}

The \emph{right to be forgotten} entitles data owners the right to delete their data from an entity storing it.
Recently enacted legislation, such as the General Data Protection Regulation (GDPR)~\cite{GDPR} in the European Union, 
the California Consumer Privacy Act (CCPA)~\cite{CCPA} in California, 
and the Personal Information Protection and Electronic Documents Act (PIPEDA)~\cite{PIPEDA} in Canada, have legally solidified this right.
Google Search has received nearly 3.2 million requests to delist certain URLs in search results over five years~\cite{BBCCCFGHHIDKLNNOPSTV19}.

In the machine learning context, the right to be forgotten requires that, in addition to the data itself, any influence of the data on the model disappears~\cite{CY15, VKL18}.
This process, also called \emph{machine unlearning}, has gained momentum both in academia and industry~\cite{CY15,XMCAR16,CYASMY18,DCLOS19,GGVZ19,SSWM20,GAS20,GGHM20,LMLLJMYR20,BSZ20, ISCZ21,NRS21,BCCJTZLP21}. 
The most legitimate way to implement machine unlearning is to remove the data sample requested to be deleted (referred to as \target sample), and retrain the ML model from scratch, but this incurs high computational overhead.
To mitigate this, several approximate approaches have been proposed~\cite{ISCZ21, CY15, BSZ20, BCCJTZLP21}.

Machine unlearning naturally generates two versions of ML models, namely the \emph{original model} and the \emph{\unlearned model}, and creates a discrepancy between them due to the \target sample's deletion.
While originally designed to protect the \target sample's privacy, we argue that machine unlearning may leave some imprint of it, and thus create unintended privacy risks.
Specifically, while the original model may not reveal much private information about the \target sample, additional information might be leaked through the \unlearned model. 

\mypara{Our Contributions}
In this paper, we study to what extent data is indelibly imprinted in an ML model by quantifying the additional information leakage caused by machine unlearning.
We concentrate on machine learning classification, the most common machine learning task, and assume both original and \unlearned models to be black-box, the most challenging setting for an adversary.

We first propose a novel membership inference attack in the machine unlearning setting that aims at determining whether the \target sample is part of the training set of the original model.
Different from \mlleak attacks~\cite{SSSS17,SZHBFB19} which leverage the output (posteriors) of a single target model,
our attack leverages outputs of both original and \unlearned models.
More concretely, we propose several aggregation methods to jointly use the two posteriors from the two models as our attack model's input, either by concatenating them or by computing their differences.
Our empirical results show that the concatenation-based methods perform better in overfitted models, while the difference-based methods perform better in well-generalized models.

Second, in order to quantify the unintended privacy risks incurred by machine unlearning, we propose two novel privacy metrics, namely \emph{\DegCount} and \emph{\DegRate}.
Both of them quantify how much relative privacy the \target has lost due to machine unlearning.
Concretely, \DegCount calculates the proportion of cases for which the adversary's confidence about the membership status of the \target sample is larger with \method than with \mlleak attack.
\DegRate calculates the average confidence increase between \method and \mlleak.

We conduct extensive experiments to evaluate the performance of \method over a series of ML models, ranging from logistic regression to convolutional neural networks, with multiple categorical datasets and image datasets.
The experimental results show that \method consistently degrades the membership privacy of the \target sample, which indicates machine unlearning can have counterproductive effects on privacy.
In particular, we observe that privacy is especially degraded because of machine unlearning in the case of well-generalized models.
For example, we observe that the \mlleak attack has an accuracy (measured by AUC) close to $0.5$, or random guessing, on the well-generalized decision tree classifier.
On the contrary, the AUC of \method is $0.89$, and the \DegCount and \DegRate are $0.85$ and $0.28$, respectively, which demonstrates that machine unlearning can have a detrimental effect on membership privacy even with well-generalized models.
We further show that we can effectively infer membership information in more practical scenarios, including the scenario where there are multiple intermediate unlearned models, the scenario where a group of samples (instead of a single one) are deleted together from the original target model, and the online learning scenario where there are samples to be deleted and added simultaneously.

Finally, in order to mitigate the privacy risks stemming from machine unlearning, we propose four possible defense mechanisms: 
(1) publishing only the top-$k$ confidence values of the posterior vector, 
(2) publishing only the predicted label,
(3) temperature scaling, and
(4) differential privacy.
The experimental results show that \method is robust to the top-$k$ defense, even when the model owner only releases the top-$1$ confidence value.
On the other hand, publishing only the predicted label, temperature scaling, and differential privacy can effectively prevent \method.

To summarize, we show that machine unlearning degrades the privacy of the \target sample in general. 
This discovery sheds light on the risks of implementing the right to be forgotten in the ML context. We believe that our attack and metrics can help develop more privacy-preserving machine unlearning approaches in the future. 
The main contributions of this paper are four-fold:
\begin{itemize}
    \item We take the first step to quantify the unintended privacy risks in machine unlearning through the lens of membership inference attacks.

    \item We propose several practical approaches for aggregating the information returned by the two versions of the ML models.

    \item We propose two novel metrics to measure the privacy degradation stemming from machine unlearning and conduct extensive experiments to show the effectiveness of \method.

    \item We propose four defense mechanisms to mitigate the privacy risks stemming from \method and empirically evaluate their effectiveness.
\end{itemize}

\mypara{Roadmap}
In \autoref{sec:preliminary}, we introduce some background knowledge about machine learning and machine unlearning, and the threat model.
\autoref{sec:attack} presents the details of our proposed attack.
We propose two privacy degradation metrics in
\autoref{sec:metric}.
We conduct extensive experiments to illustrate the effectiveness of the proposed attack in
\autoref{sec:exp} and \autoref{sec:practical_deployment}.
In \autoref{sec:defense}, we introduce several possible defense mechanisms and empirically evaluate their effectiveness.
We discuss the related work in \autoref{sec:related} and conclude the paper in \autoref{sec:conclusion}.

\section{Preliminaries}
\label{sec:preliminary}

\subsection{Machine Learning}

In this paper, we focus on machine learning classification, the most common ML task.
An ML classifier \model{}{} maps a data sample $x$ to posterior probabilities \post{}{},
where \post{}{} is a vector of entries indicating the probability of $x$ belonging to a specific class $y$ according to the model $\model{}{}$. 
The sum of all values in \post{}{} is $1$ by definition. 
To construct an ML model, one needs to collect a set of data samples, referred to as the training set \dset{}{}.
The model is then built through a training process that aims at minimizing a predefined loss function with some optimization algorithms, such as stochastic gradient descent.

\subsection{Machine Unlearning}

Recent legislation such as GDPR and CCPA enact the ``right to be forgotten'', which allows individuals to request the deletion of their data by the model owner to preserve their privacy.
In the context of machine learning, e.g., MLaaS, this implies that the model owner should remove the \textit{target sample} $x$ from its training set \dset{}{}.  
Moreover, any influence of $x$ on the model \model{}{} should also be removed.
This process is referred to as machine unlearning. 

\mypara{Retraining from Scratch}
The most legitimate way to implement machine unlearning is to retrain the whole ML model from scratch.
Formally, denoting the \textit{original model} as \model{o}{} and its training dataset as \dset{o}{},
this approach consists of training a new model \model{u}{} on dataset $\dset{u}{} = \dset{o}{} \setminus x$.\footnote{Note that we also study the removal of more than one sample in our experimental evaluation, but for simplicity we formalize our problem with one sample only.}
We call this \model{u}{} the \textit{unlearned model}.

Retraining from scratch is easy to implement.
However, when the size of the original dataset \dset{o}{} is large and the model is complex, the computational overhead of retraining is too large. 
To reduce the computational overhead,
several approximate approaches have been proposed~\cite{ISCZ21, CY15, BSZ20, BCCJTZLP21}.

\mypara{SISA}
\sisa~\cite{BCCJTZLP21} works in an ensemble style, which is an efficient and general method to implement machine unlearning.
The training dataset \dset{o}{} in \sisa is partitioned into $k$ disjoint parts $\dset{o}{1}, \dset{o}{2}, \cdots, \dset{o}{k}$.
The model owner trains a set of original ML models $\model{o}{1}, \model{o}{2}, \cdots, \model{o}{k}$ on each corresponding dataset $\dset{o}{i}$.
When the model owner receives a request to delete a data sample $x$, it just needs to retrain the sub-model $\model{o}{i}$ that contains $x$, resulting in unlearned model $\model{u}{i}$.
Sub-models that do not contain $x$ remain unchanged.
Notice that the size of dataset $\dset{o}{i}$ is much smaller than $\dset{o}{}$; thus, the computational overhead of \sisa is much smaller than the ``retraining from scratch'' method.

At inference time, the model owner aggregates predictions from the different sub-models to provide an overall prediction.
The most commonly used aggregation strategy is majority vote and posterior average.
In our experiments, we use posterior average as aggregation strategy.

\subsection{Threat Model}
\label{subsec:threat_model}

\mypara{Adversary's Goal}
Given a target sample $x$, an original model, and its unlearned model, the adversary aims to infer whether $x$ is unlearned from the original model.
In other words, the adversary aims to know that the target sample is in the training dataset of the original model but it is not in the training dataset of the unlearned model.
While the goal of unlearning $x$ is to protect $x$'s privacy, a successful attack considered here can show unlearning instead jeopardizes $x$'s privacy (especially when $x$'s membership leakage risk is not severe on the original model before machine unlearning).

We focus on membership privacy as it is one of the most established method on quantifying privacy risks of ML models~\cite{SSSS17}.
Knowing that a specific data sample $x$ was used to train a particular model may lead to potential privacy breaches. 
For example, knowing that a certain patient's clinical records were used to train a model associated with a disease 
(e.g., to determine the appropriate drug dosage or to discover the genetic basis of the disease) 
can reveal that the patient carries the associated disease.
Unlike classical membership inference, which only leverages the output of a target ML model, our adversary can exploit information of both original and unlearned models to perform their attack.

\mypara{Other Cases}
Besides the samples that are in the original model's training dataset but not in the unlearned model's training dataset, referred to as \tuple{in, out}, there are other three cases, including \tuple{out, out}, \tuple{in, in}, \tuple{out, in}.
Samples in the \tuple{out, out} category are considered as the negative cases for the adversary to train their attack model (see \autoref{subsec:attack_model_train}).
For the \tuple{in, in} samples, their membership privacy can be quantified by a classical membership inference attack. The difference is that the attack model can leverage both original and unlearned models' information (see \autoref{subsec:non_redacted_samples}).
It is also possible that during the unlearning process, the unlearned model unlearns samples from the target model and updates itself with other new samples (referred to as incremental learning).
In such cases, we have \tuple{out, in} samples.
To infer the membership status of such samples, the adversary can similarly perform a classical membership inference attack on the unlearned model.
Note that the privacy of \tuple{out, in} samples are not directly related to the privacy risks caused by machine unlearning, and the current literature on machine unlearning~\cite{BCCJTZLP21} also does not consider such cases.
In \autoref{subsec:online_learning}, we evaluate our attack when unlearning and updating are both performed on the target model simultaneously. 

\mypara{Adversary's Knowledge}
We assume that the adversary has black-box access to an original ML model and its unlearned model.
This is realistic as the target black-box model can be queried at any time, such as in the setting of MLaaS, and all of the query results can be stored locally by the adversary; this also follows the assumption of previous works~\cite{SBBFZ20,SSSS17,SZHBFB19,BWTRPOKB20,SS20}.
As such, when there are no changes on the target sample's outputs from two consecutive queries, then the unlearning did not happen and the adversary does not need to launch the attack.
On the other hand, when the adversary observes changes on the target sample's outputs, they know that the target model has been updated.
In most parts of the paper, we consider the scenario where the original model and the unlearned model differ only by one sample.
However, we further show that when the two models differ by multiple samples, the adversary can still mount their attack effectively (see~\autoref{sec:practical_deployment}).

We also assume that the adversary has a local shadow dataset which can be used to train a set of shadow models to mimic the behavior of the target model. 
The shadow models are then used to generate training data for the attack model (see \autoref{sec:attack} for more details).
The shadow dataset can either come from the same distribution as the target dataset or from a different one. 
We evaluate both settings in \autoref{sec:exp}.

\mypara{Difference with Updates-Leak~\cite{SBBFZ20}}
There are recent studies aiming to quantify the information leakage in the model updating process.
Salem et al.~\cite{SBBFZ20} show that in the online learning applications, where an ML image classifier is updated by new data samples, the adversary can reconstruct the updated samples by exploiting information from two versions of the target ML model (before and after the updating).
Brockschmidt et al.~\cite{BWTRPOKB20} show similar results in the natural language models as well as the data deletion scenario.
This line of work is related to our attack in the sense that we all study the unintended information leakage in model updating processes.
However, the attack goals are different.

Our attack focuses on membership inference while Salem et al.~\cite{SBBFZ20} propose two attacks (in the single-sample setting), including single-sample reconstruction and single-sample label inference.
The former (single-sample reconstruction) can only reconstruct a data sample that is ``similar'' to the updated sample; 
but it cannot further determine the membership status of the target sample by simply comparing the difference between the reconstructed sample and the target sample, and concluding the membership status with a predefined threshold.
One reason is that it is unclear which metric is the best to measure two samples' similarity.
Updates-Leak uses MSE, which is not ideal.
A possible way to address this is to involve humans in the loop to judge samples' similarities visually.
Nevertheless, besides the scalability issues, the qualitative results presented in~\cite{SBBFZ20} show that the reconstructed samples are not very visually close to the original ones (see Figure 8 of~\cite{SBBFZ20}).
Another reason is that Updates-Leak relies on a pretrained autoencoder which may introduce biases to the reconstructed sample.
The latter (single-sample label inference) is a coarser-grained attack aiming at ``class-level'' inference, while the membership inference attack in this paper is a finer-grained attack aiming at ``sample-level'' inference.
We emphasize that membership inference is the most well-established privacy attack and arguably constitutes a bigger privacy threat~\cite{SRS17,SZHBFB19,MSCS19,NSH19,HMDC19,CYZF20,SS19,HRSF20,YGFJ18,LF20} than class-level attacks, such as the attacks in~\cite{SBBFZ20,BWTRPOKB20} and model inversion~\cite{FLJLPR14,FJR15}.

\section{Membership Inference in Machine Unlearning}
\label{sec:attack}

\begin{figure}[!t]
\begin{center}
\includegraphics[width=0.47\textwidth]{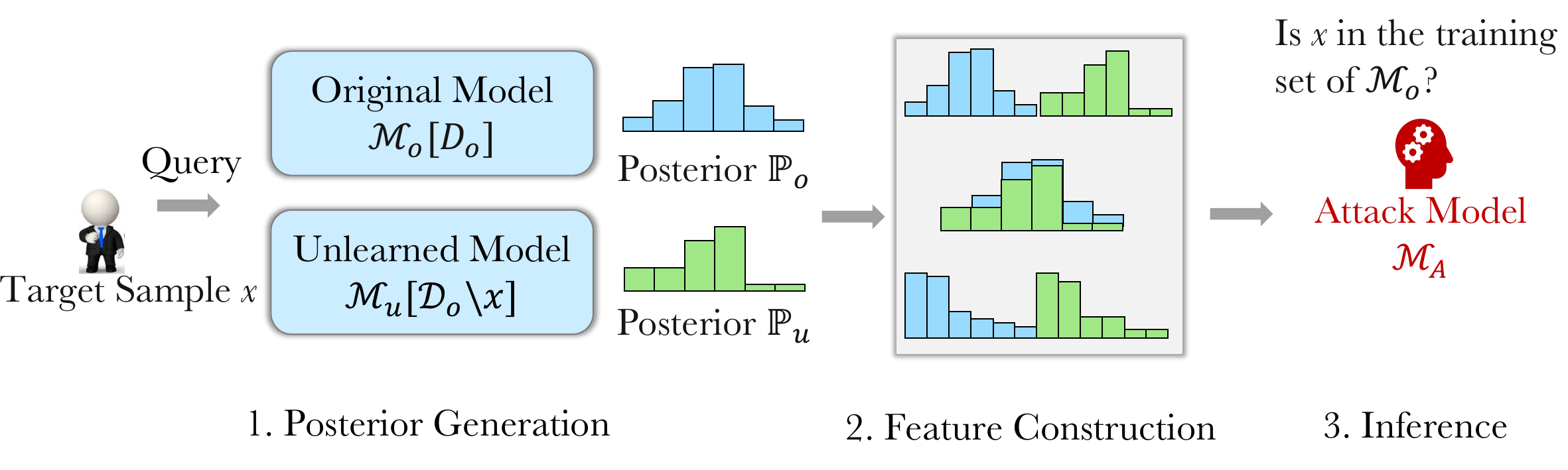}
\end{center}
\vspace{-0.5cm}
\caption{A schematic view of the general attack pipeline. 
The membership status of the target sample $x$ is leaked due to the two versions of model.} 
\label{fig:pipeline}
\end{figure}

\subsection{Attack Pipeline}
\label{subsec:pipeline}
The general attack pipeline of \method is illustrated in \autoref{fig:pipeline}.  
It consists of three phases:
posteriors generation, feature construction and (membership) inference.

\mypara{Posteriors Generation}
The adversary has access to two versions of the target ML model, the original model $\model{o}{}$ and the \unlearned model $\model{u}{}$.
Given a \target sample $x$, the adversary queries $\model{o}{}$ and $\model{u}{}$, and obtains the corresponding posteriors, i.e., \post{o}{} and \post{u}{}, also referred to as confidence values~\cite{SSSS17}.

\mypara{Feature Construction}
Given the two posteriors \post{o}{} and \post{u}{}, the adversary aggregates them to construct the feature vector \feat{}.
There are several alternatives to construct the feature vector.
We discuss five representative methods in \autoref{subsec:feature}.

\mypara{Inference}
Finally, the adversary sends the obtained \feat{} to the attack model, which is a binary classifier, to determine whether the \target sample $x$ is in the training set of the original model.
We describe how to build the attack model in \autoref{subsec:attack_model_train}.

\begin{figure*}[!ht]
\begin{center}
\includegraphics[width=0.9\textwidth]{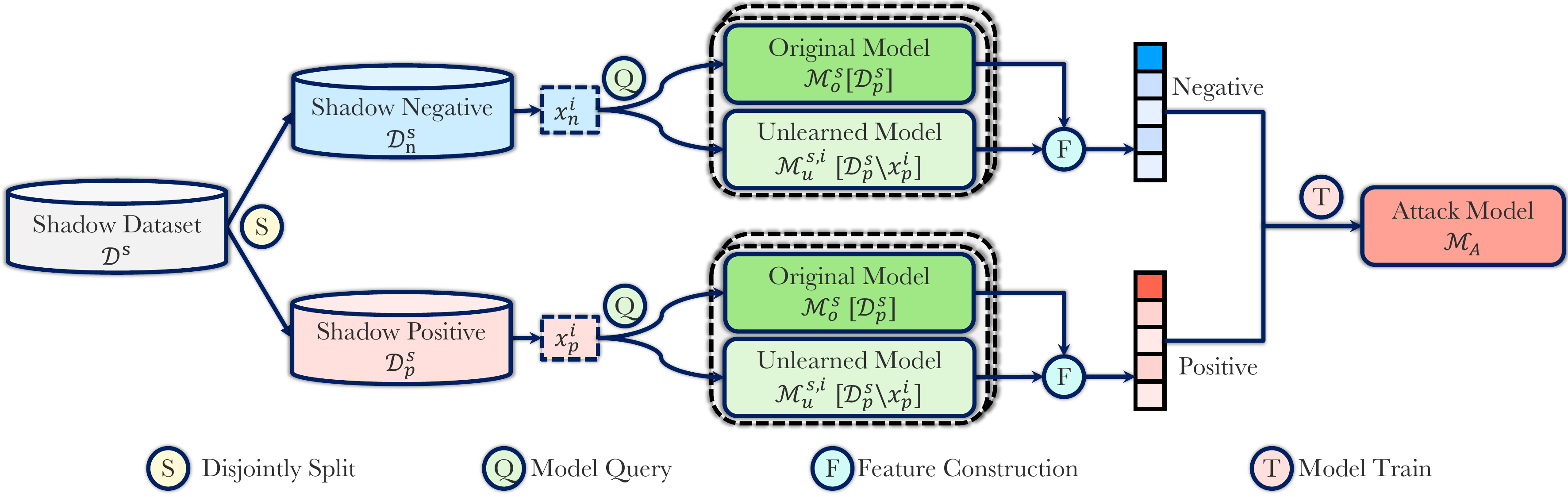}
\end{center}
\vspace{-0.5cm}
\caption{Training process of the attack model. 
The shadow dataset \dset{}{s} is split into disjoint shadow positive dataset \dset{p}{s} and shadow negative dataset \dset{n}{s}.
The shadow positive dataset \dset{p}{s} is used to train the shadow original model \model{o}{s}.
The shadow \unlearned model \model{u}{s,i} is trained on $\dset{p}{s} \setminus x_p^i$, where $x_p^i \in \dset{p}{s}$.
In the inference phase, the adversary first uses \target sample $x_p^i$ to query the original and \unlearned models simultaneously to generate the positive features.
Then they use a random sample $x_n^i \in \dset{n}{s}$ to query the corresponding models to generate the negative features.
Finally, they use the positive and negative features to train the attack model \model{A}{}.
}
\label{fig:attack_model_train}
\end{figure*}

\subsection{Attack Model Training}
\label{subsec:attack_model_train}
We assume the adversary has a local dataset, which we call the \textit{shadow dataset} \dset{}{s}. 
The shadow dataset can come from a different distribution than the one used to train the target model.
To infer whether the \target sample $x$ is in the original model or not, our core idea is to train an attack model $\model{A}{}$ that captures the difference between the two posteriors.
The intuition is that, if the \target sample $x$ is deleted, the two models $\model{o}{}$ and $\model{u}{}$ will behave differently.
\autoref{fig:attack_model_train} illustrates the training process of the attack model, and the detailed training procedure is presented as follows.

\mypara{Training Shadow Models}
To mimic the behavior of the target model, the adversary needs to train a shadow original model and a set of shadow \unlearned models.
To do this, the adversary first partitions \dset{}{s} into two disjoint parts, the shadow negative set \dset{n}{s} and the shadow positive set \dset{p}{s}.
The shadow positive set \dset{p}{s} is used to train the shadow original model \model{o}{s}.
The shadow \unlearned model \model{u}{s} is trained by deleting samples from \dset{p}{s}.
For ease of presentation, we assume the shadow \unlearned model \model{u}{s} is obtained by deleting exactly one sample.
We will show that \method is still effective for group deletion in \autoref{subsec:group_deletion}.
The adversary randomly generates a set of deletion requests (\target samples) $\mathcal{R}_p = \{x_p^1, x_p^2, \cdots, x_p^m\}$ and train a set of shadow \unlearned models $\model{u}{s,1}, \model{u}{s,2}, \cdots, \model{u}{s,m}$, where the shadow \unlearned model \model{u}{s,i} is trained on dataset $\dset{p}{s} \setminus x_p^i$.

\mypara{Obtaining Posteriors}
At the posteriors generation phase, the adversary feeds each \target sample $x_p^i \in \mathcal{R}_p$ to the shadow original model \model{o}{s} and its corresponding shadow \unlearned model \model{u}{s,i}, and gets two posteriors \post{o}{i} and \post{u}{i}.

\mypara{Constructing Features}
The adversary then uses the feature construction methods discussed in \autoref{subsec:feature} to construct training cases for the attack model.
In \mlleak, posteriors of $x_p^i \in \mathcal{R}_p$ serve as member cases of the attack model.
But in the machine unlearning setting, $x_p^i \in \mathcal{R}_p$ is member of the shadow original model \model{o}{s} and non-member of the shadow \unlearned model \model{u}{s}.
To avoid confusion, we call the samples related to $x_p^i \in \mathcal{R}_p$ \emph{positive} cases instead of member cases for the attack model.

To train the attack model, the adversary also needs a set of negative cases.
This can be done by sampling a set of negative query samples $\mathcal{R}_n$ from the shadow negative dataset \dset{n}{s} and query the shadow original model and \unlearned model.
To get a good attack model generalization performance, the adversary needs to ensure that the number of positive cases and the number of negative cases of the attack model are balanced, i.e., $|\mathcal{R}_p|=|\mathcal{R}_n|$, where $|\cdot|$ is the cardinality of the sample set. 

\mypara{Improving Diversity}
To improve the diversity of the attack model, the adversary obtains multiple shadow original models by randomly sampling multiple subsets of samples from the shadow positive dataset \dset{p}{s}.
For each shadow original model, the adversary randomly generates a set of deletion requests and trains a set of shadow \unlearned models.
In \autoref{subsec:hyperparameter}, we conduct empirical experiments to show the impact of the number of shadow original models on the attack performance.

\mypara{Training the Attack Model}
Given sets of positive cases with features and negative cases with features, we adopt four standard and widely used classifiers as our attack model: Logistic regression, decision tree, random forest, and multi-layer perceptron.

\subsection{Feature Construction}
\label{subsec:feature}
Given the two posteriors, a straightforward approach to aggregate the information is to concatenate them, i.e., $\post{o}{} || \post{u}{}$, where $||$ is the concatenation operation.  
This preserves the full information.
However, it is possible that the concatenation contains redundancy. 
In order to reduce redundancy, we can instead rely on the difference between $\post{o}{}$ and $\post{u}{}$ to capture the discrepancy left by the deletion of the \target sample.  
In particular, we make use of the element-wise difference $\post{o}{} - \post{u}{}$ and the Euclidean distance $\left\|\post{o}{} - \post{u}{}\right\|_2$.

In order to better capture the level of confidence of the model, one may also sort the posteriors before the difference or concatenation operations~\cite{GWYGB18}.  
Specifically, we sort the original posteriors \post{o}{} in descending order and get the sorted original posteriors \post{o}{s}.
We then rearrange the order of the \unlearned posteriors \post{u}{} to align its elements with \post{o}{}, and get the sorted \unlearned posteriors \post{u}{s}.

To summarize, we adopt the following five methods to construct the features for the attack model:
\begin{itemize}
    \item Direct concatenate (\dc), i.e., $\post{o}{} || \post{u}{}$
    \item Sorted concatenate (\scc), i.e., $\post{o}{s} || \post{u}{s}$
    \item Direct difference (\dd), i.e., $\post{o}{} - \post{u}{}$.
    \item Sorted difference (\sd), i.e., $\post{o}{s} - \post{u}{s}$.
    \item Euclidean distance (\ed), i.e., $\left\|\post{o}{} - \post{u}{}\right\|_2$
\end{itemize}

In \autoref{subsec:choice_feature}, we conduct empirical experiments to evaluate the performance of the above methods and provide a high-level summary of the best features to use depending on the behavior of the underlying ML model.

\section{Privacy Degradation Measurement}
\label{sec:metric}
In this paper, we aim to evaluate to what extent machine unlearning may degrade the membership privacy of an individual whose data sample has been deleted from the training set (we also call this the \target sample). 
Specifically, we want to quantify the additional privacy degradation \method brings over \mlleak (or the improvement of membership inference) in order to measure the unintended information leakage due to data deletion in machine learning.
To this end, we propose two privacy degradation metrics that measure the difference in the confidence levels of \method and \mlleak when predicting the correct  membership status of the \target sample.

Given $n$ \target samples $x^1$ to $x^n$, define $p_u^i$ as the confidence of \method in classifying $x^i$ as a member, and $p_m^i$ as the confidence of \mlleak.  
Let $b^i$ be the true status of $x^i$, i.e., $b^i=1$ if $x^i$ is a member, and $b^i=0$ otherwise.
With that, we define the following two metrics:

\begin{itemize}
    \item \textbf{\betrate.}
    \betrate stands for \DegCount.  It calculates the proportion of \target samples whose true membership status is predicted with higher confidence by \method than by \mlleak.
    Formally, \betrate is defined as
    $$
    \betrate = \frac{1}{n}\sum_i^n \left[ b^i\mathbbm{1}_{p_u^i > p_m^i} + (1-b^i)\mathbbm{1}_{p_u^i < p_m^i}\right]
    $$
    where $\mathbbm{1}_{P}$ is the indicator function which equals $1$ if $P$ is true, and $0$ otherwise.
    Higher \betrate means higher privacy degradation.
    
    \item \textbf{\imprate.}
    \imprate stands for \DegRate.  It calculates the average confidence improvement rate of \method predicting the true membership status compared to \mlleak.
    \imprate can be formally defined as
    $$
    \imprate =\frac{1}{n}\sum_i^n  \left[b^i(p_u^i - p_m^i) + (1 - b^i)(p_m^i - p_u^i)\right]
    $$
    Higher \imprate means higher privacy degradation.
\end{itemize}

\section{Evaluation}
\label{sec:exp}

In this section, we conduct extensive experiments to evaluate the unintended privacy risks of machine unlearning.
We first conduct an end-to-end experiment to validate the effectiveness of \method on multiple datasets using the most straightforward unlearning method, i.e., retraining from scratch.
Second, we compare different feature construction methods proposed in \autoref{subsec:feature} and provide a summary of the most appropriate to choose depending on the context.
Third, we evaluate the impact of overfitting and of different hyperparameters.
Fourth, we conduct experiments to evaluate dataset and model transferability between shadow model and target model.
Finally, we show the effectiveness of our attack against the \sisa unlearning method.
We leave the evaluation of our attack in other scenarios to~\autoref{sec:practical_deployment}.

\subsection{Experimental Setup}
\label{subsec:experiment_setup}

\mypara{Target Models}
In our experiments, we evaluate the vulnerability of both simple machine learning models, including logistic regression (LR), decision tree (DT), random forest (RF), and 5-layer multi-layer perceptron (MLP), and the state-of-the-art convolutional neural networks, including SimpleCNN (implemented by us), DenseNet~\cite{HLMW17}, and ResNet50~\cite{HZRS16}.
For reproducibility purpose, we provide the hyperparameter settings for the simple machine learning models in \autoref{app:model_setting}, and the implementation details of SimpleCNN in \autoref{app:cnn_structure}.
All the convolutional networks are trained for 100 epochs.

\begin{table}[!t]
    \centering
    \caption{Dataset statistics.}
    \label{table:dataset}
    \resizebox{1.0\linewidth}{!}{
    \setlength{\tabcolsep}{0.12em}
    \renewcommand{\arraystretch}{1.3}
    \begin{tabular}{c c c c c}
    \toprule
    \textbf{Dataset} & \textbf{Type} & \textbf{Feature Dimension} & \textbf{\#. Classes} & \textbf{\#. Samples} \\ 
    \toprule
    Adult & Categorical & 14 & 2 & 50,000 \\
    \rowcolor{mygray}
    Accident & Categorical & 30 & 3 & 3,000,000 \\
    Insta-NY & Categorical & 169 & 9 & 19,215 \\
    \rowcolor{mygray}
    Insta-LA & Categorical & 169 & 9 & 16,472 \\
    MNIST & Image & 28*28*1 & 10 & 42,000 \\
    \rowcolor{mygray}
    CIFAR10 & Image & 32*32*3 & 10 & 60,000 \\
    STL10 & Image & 32*32*3 & 10 & 13,000 \\
    \bottomrule
    \end{tabular}
    }
\end{table}

\mypara{Datasets}
We run experiments on two different types of datasets: categorical datasets and image datasets.
The categorical datasets are used to evaluate the vulnerability of simple machine learning models, while the image datasets are used to evaluate the vulnerability of the convolutional neural networks.
Due to space limitation, we defer the detailed description of these datasets to \autoref{app:dataset}.
The statistics of all datasets used in our experiments are listed in \autoref{table:dataset}.
Note that the Insta-NY and STL10 datasets are only used for data transferring attack in \autoref{subsec:transfer_attack}.

\begin{figure*}[!t]
    \centering
    \includegraphics[width=1\textwidth]{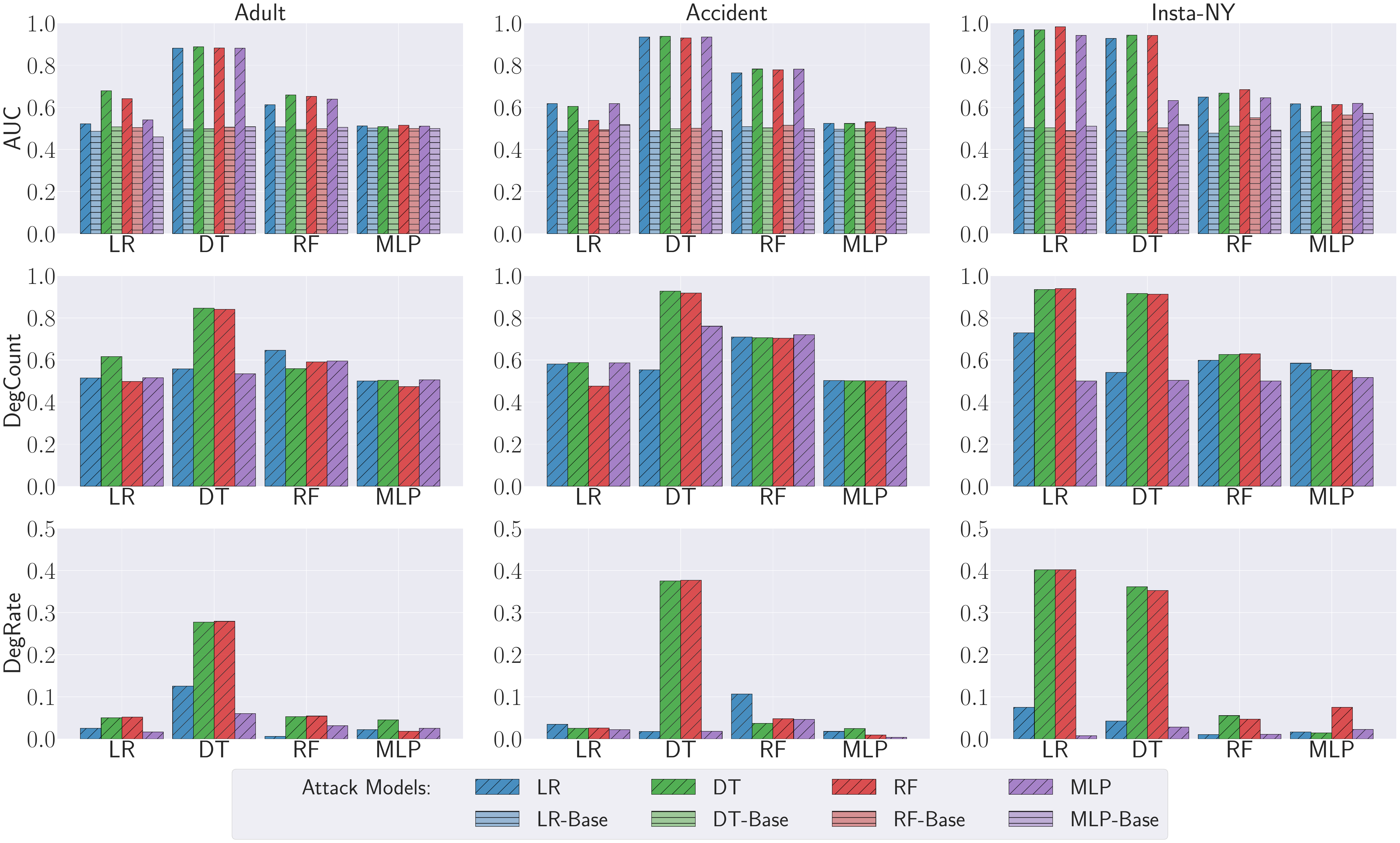}
    \caption{
    Privacy degradation level on the \scratch method for three categorical datasets.
    Rows stand for different metrics, columns stand for different datasets.
    In each subfig, the groups in x-axis represent different target models, and the legends (in different colors) represent different attack models.
    For the AUC metric, the right bars (transparent ones) stand for the AUC value of \mlleak.
    }
    \label{fig:scratch_compare}
\end{figure*}

\mypara{Metrics}
In addition to the two privacy degradation metrics proposed in \autoref{sec:metric}, we also rely on the traditional AUC metric to measure the absolute performance of \method and \mlleak.
To summarize, we have the following three metrics:

\begin{itemize}
    \item \textbf{AUC.}
    It is a widely used metric to measure the performance of binary classification in a range of thresholds~\cite{FLJLPR14,BHPZ17,PTC18,HZHBTWB19,SZHBFB19,JSBZG19}. 
    An AUC value equals to 1 shows a maximum performance while an AUC value of 0.5 shows a performance equivalent to random guessing. 
    \item \textbf{\betrate.}
    It stands for \DegCount, which is defined in \autoref{sec:metric}.
    \item \textbf{\imprate.}
    It stands for \DegRate, which is defined in \autoref{sec:metric}.
\end{itemize}

\mypara{Experimental Settings}
We evenly split each dataset \dset{}{} into disjoint target dataset \dset{}{t} and shadow dataset \dset{}{s}.
In \autoref{subsec:transfer_attack}, we will show that the shadow dataset can come from a different distribution than the target dataset.
The shadow dataset \dset{}{s} is further split into shadow positive dataset \dset{p}{s} and shadow negative dataset \dset{n}{s} ($80\%$ for \dset{p}{s} and $20\%$ for \dset{n}{s}).
We randomly sample \so subsets of samples from \dset{p}{s}, each containing \sr samples, to train \so shadow original models.
For each shadow original model \model{o}{s,i}, we train \su shadow \unlearned models on $\dset{o}{s,i} \setminus x$.
We split the target dataset \dset{}{t} in a similar way as the shadow dataset \dset{}{s}.

By default, we set the hyperparameters of the shadow models to $\so=20, \sr=5000, \su=100$, and the corresponding hyperparameters of the target models to $\too=20, \tr=5000, \tu=100$.
These hyperparameters have shown to achieve good balance between computational overhead and attack performance in \autoref{subsec:hyperparameter}.

\mypara{Implementation}
All algorithms are implemented in Python 3.7 and the experiments are conducted on a Ubuntu 19.10 LTS server with Intel Xeon E7-8867 v3 @ 2.50GHz and 1.5TB memory.

\begin{figure*}[!t]
\centering
    \begin{minipage}{1\textwidth}
    \begin{subfigure}{0.33\columnwidth}
    \includegraphics[width=\columnwidth]{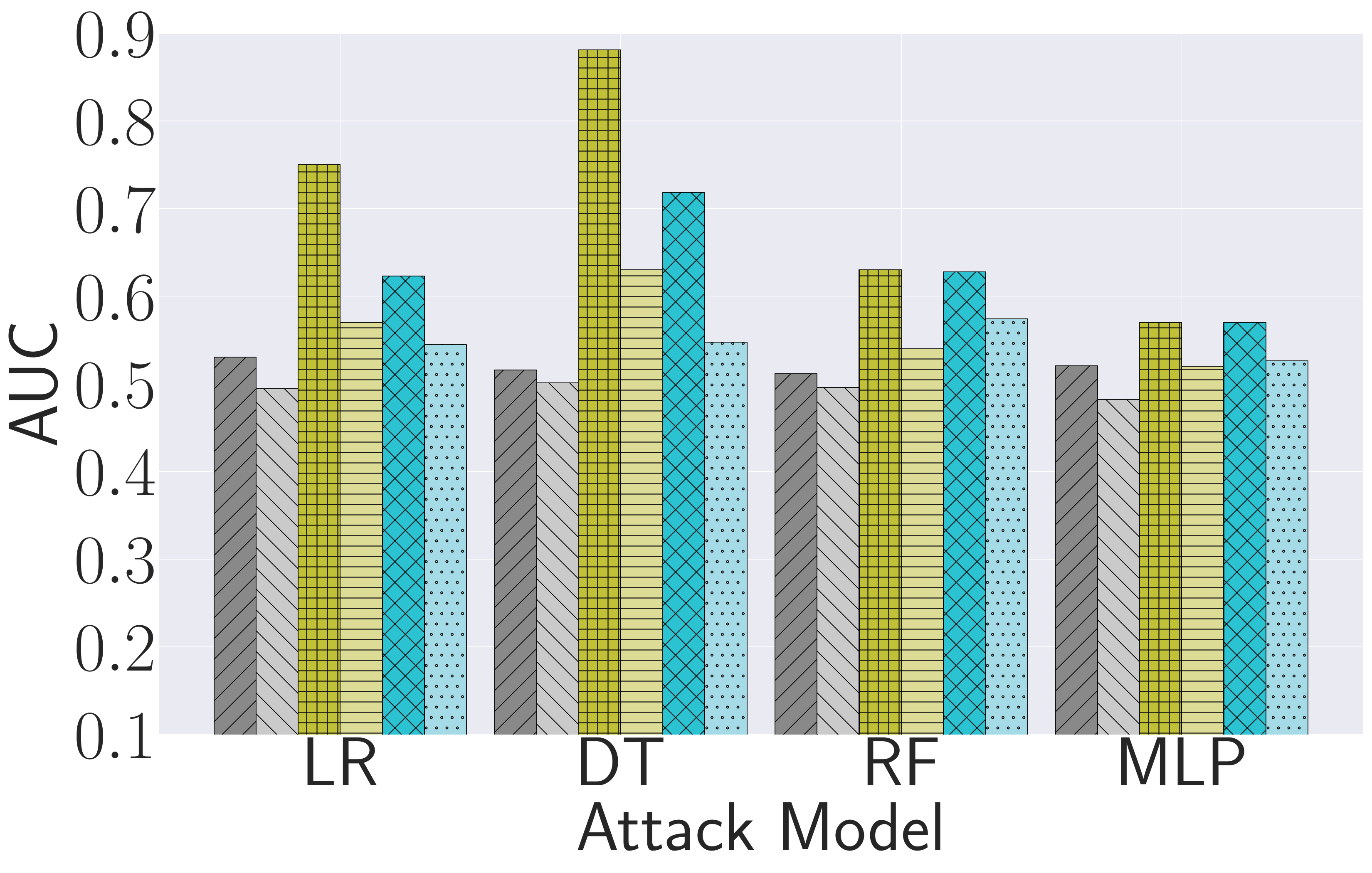}
    \label{subfig:scratch_compare_image_auc}
    \end{subfigure}
    \begin{subfigure}{0.33\columnwidth}
    \includegraphics[width=\columnwidth]{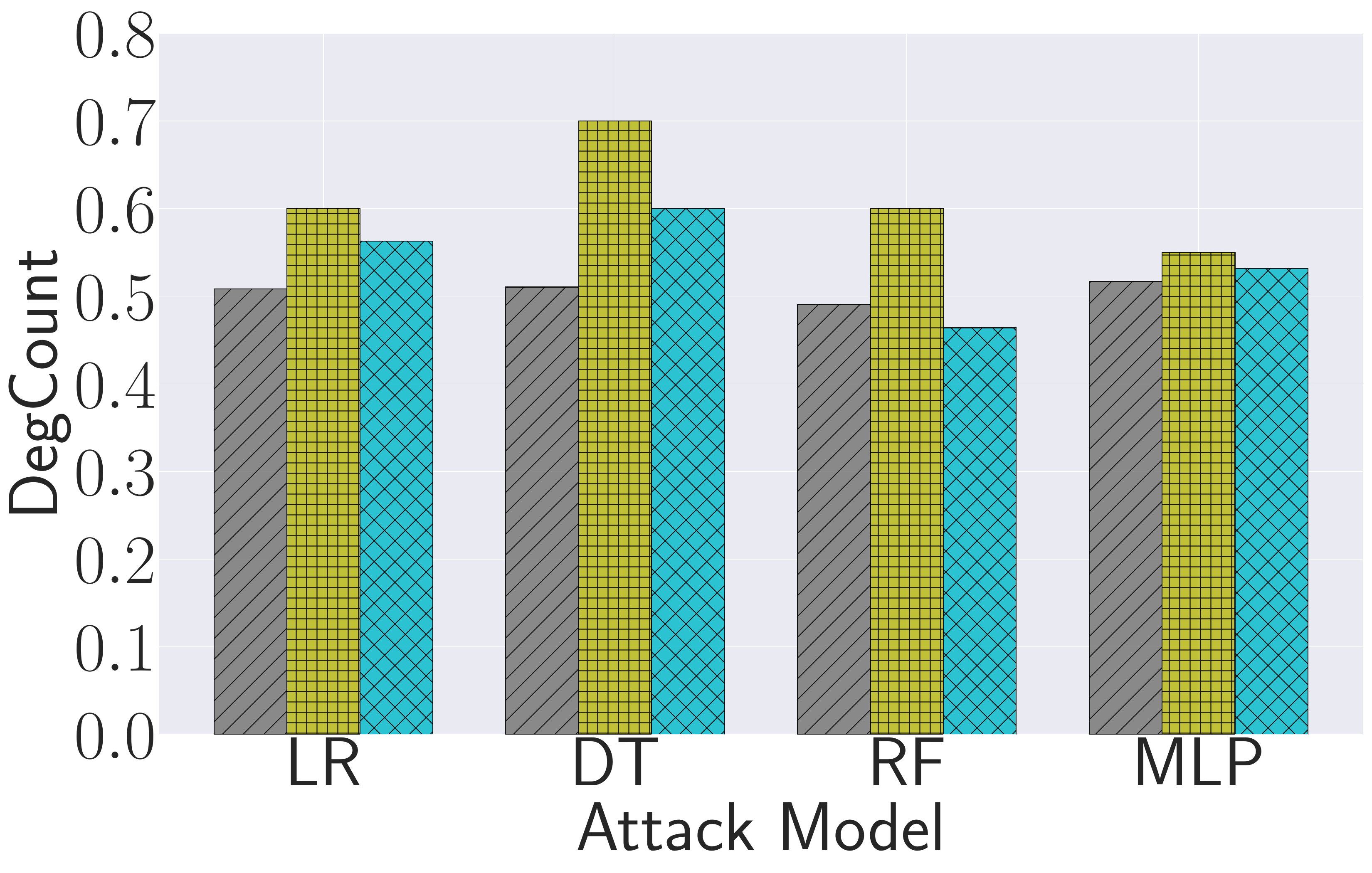}
    \label{subfig:scratch_compare_image_count}
    \end{subfigure}
    \begin{subfigure}{0.33\columnwidth}
    \includegraphics[width=\columnwidth]{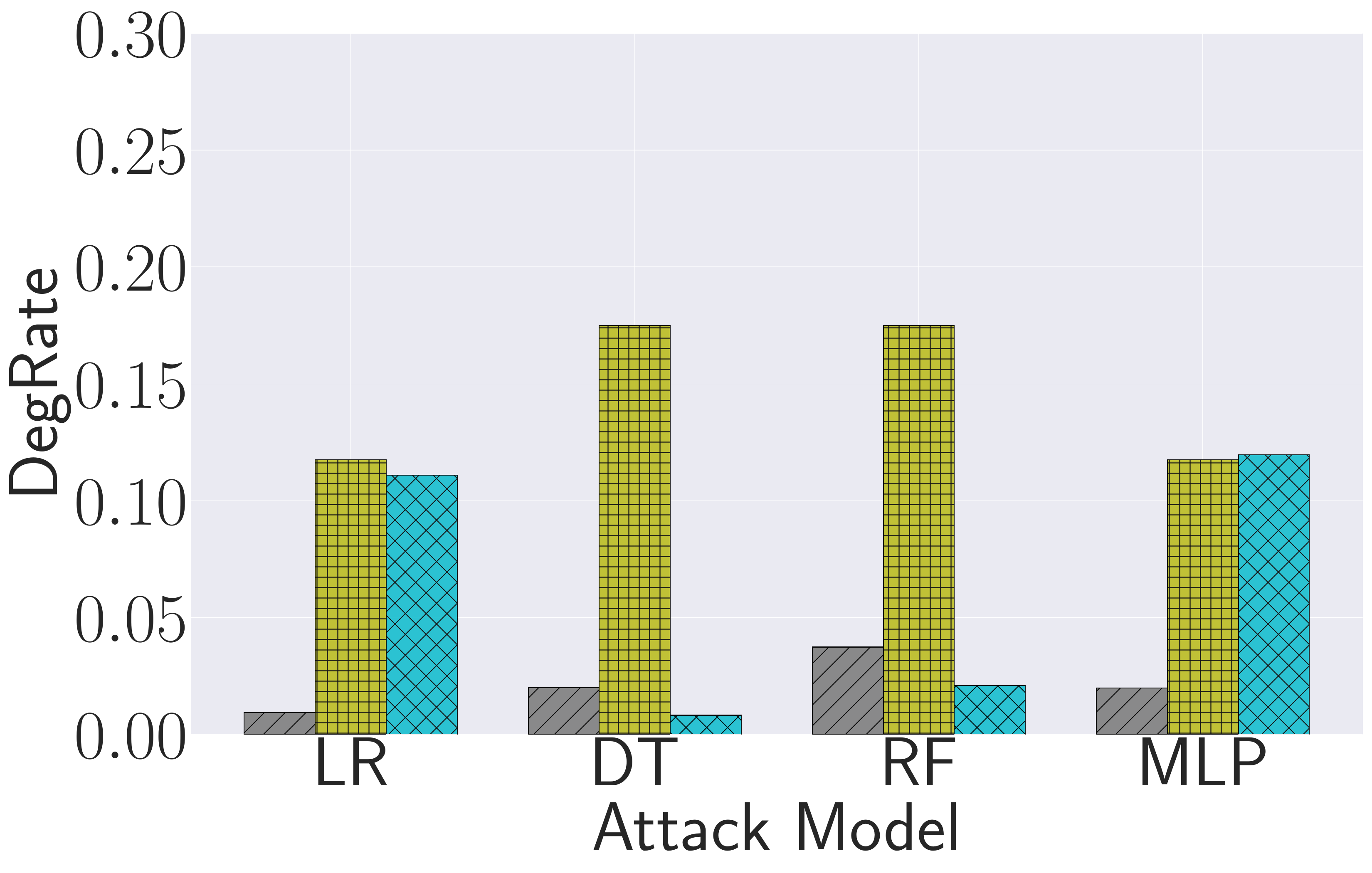}
    \label{subfig:scratch_compare_image_rate}
    \end{subfigure} \\ [-4ex]
    \medskip
    \end{minipage}
    \subfloat{\includegraphics[width=0.8\textwidth]{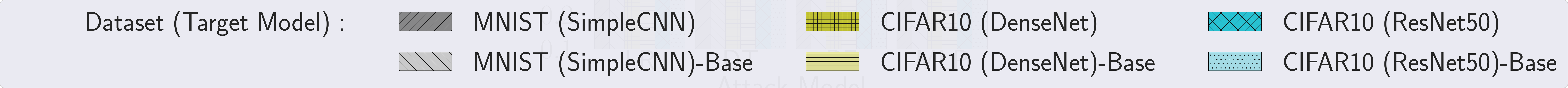}} \\
    \caption{
    Privacy degradation level on the \scratch method for image datasets.
    In each subfig, the groups in x-axis stand for different attack models, the legends (in different colors) stand for different datasets and the corresponding target models.
    For the AUC metric, the right bars (transparent ones) stand for the AUC value of \mlleak.
    }
    \label{fig:scratch_compare_image}
\end{figure*}

\subsection{Evaluation of the \scratch Method}

In this subsection, we conduct end-to-end experiments to evaluate our attack against the most straightforward approach of retraining the ML model from scratch.

\mypara{Setup}
We start by considering the scenario where only one sample is deleted for each \unlearned model.
The scenario where multiple samples are deleted before the ML model is retrained will be evaluated in \autoref{subsec:group_deletion}.
We conduct the experiment on both categorical datasets and image datasets with three evaluation metrics, namely AUC, \betrate, \imprate, and report the results with the optimal features as explained in \autoref{subsec:choice_feature}.

\mypara{Results for Categorical Datasets}
\autoref{fig:scratch_compare} depicts the attack performance of categorical datasets.
In general, we observe that \method performs consistently better than \mlleak on all datasets, target models, attack models, and metrics.
Compared to \mlleak, \method achieves up to $0.48$ improvement of the AUC.
The best \betrate and \imprate values are of $0.94$ and $0.40$, respectively.
This indicates that \method indeed degrades membership privacy of the target sample in the machine unlearning setting.
Comparing the performance of different target models, we observe that decision tree is the most vulnerable ML model.
We posit this is due to the fact that the decision tree forms a tree structure and deleting one sample could explicitly change its structure; thus the posterior difference of decision tree's original model and unlearned model is more significant, leading to a better attack performance.

\mypara{Results for Image Datasets}
\autoref{fig:scratch_compare_image} illustrates the performance for the image datasets and complex convolutional neural networks.
We keep the same attack models as categorical datasets, and use the SimpleCNN model for MNIST, use the ResNet50 and DenseNet models for CIFAR10.
In general, we also observe that our attack outperforms classic membership inference attack in all settings.
Besides, CIFAR10 trained with DenseNet shows the highest privacy degradation, while MNIST dataset trained with SimpleCNN shows the lowest.
The reason behind is that the overfitting level of CIFAR10 trained with DenseNet is the largest.
To further confirm this, we list the overfitting level of different models in
\autoref{table:overfitting}.
We observe that the overfitting level of CIFAR10 trained with DenseNet is $0.439$, while the MNIST dataset trained with SimpleCNN has an overfitting level smaller than $0.05$.

\subsection{Finding Optimal Features}
\label{subsec:choice_feature}
\begin{figure}[!tpb]
    \centering
    \begin{subfigure}{0.7\columnwidth}
    \includegraphics[width=\textwidth]{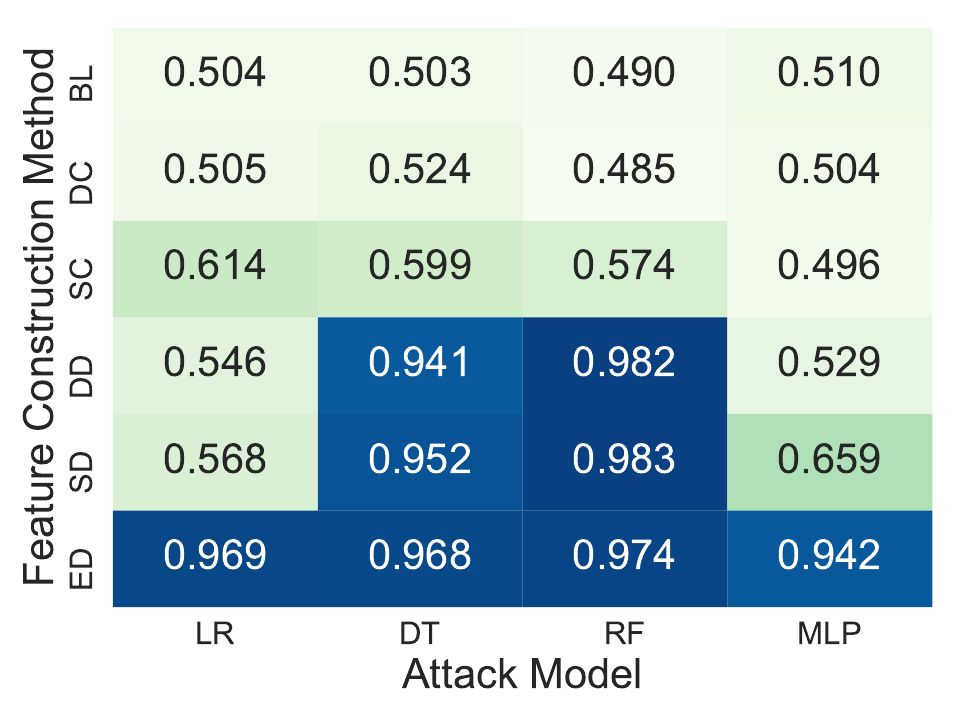}
    \vspace{-0.6cm}
    \caption{LR + Insta-NY}
    \label{subfig:feature_lr}
    \end{subfigure} \\ [+1ex]
    \begin{subfigure}{0.7\columnwidth}
    \includegraphics[width=\textwidth]{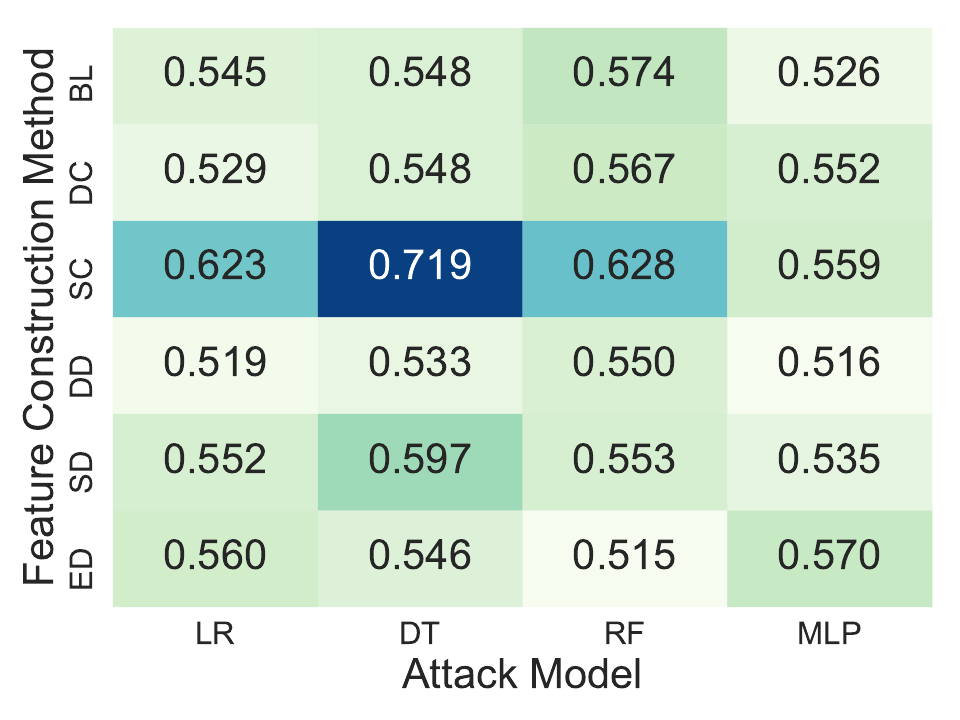} 
    \vspace{-0.6cm}
    \caption{ResNet50 + CIFAR10}
    \label{subfig:feature_rf}
    \end{subfigure} \\ 
    \caption{Attack AUC for different feature construction methods for target models (a) logistic regression (trained on Insta-NY) and (b) ResNet50 (trained on CIFAR10).
    $\mathsf{DC}$, $\mathsf{SC}$, $\mathsf{DD}$, $\mathsf{SD}$, $\mathsf{ED}$ stand for \dc, \scc, \dd, \sd, \ed, respectively.
    $\mathsf{BL}$ stands for the baseline, i.e., \mlleak.}
    \label{fig:feature_constructin_compare}
\end{figure}

\autoref{fig:feature_constructin_compare} illustrates the attack AUC of different feature construction methods.
We compare two different types of target models: (a) the well-generalized model logistic regression (trained on Insta-NY dataset), and (b) the overfitted model ResNet50 (trained on CIFAR10 dataset).
We then apply the $5$ different feature construction methods proposed in \autoref{subsec:feature} to $4$ different attack models, resulting in $20$ combinations.
For comparison, we also include the \mlleak as a baseline.

\mypara{Concatenation vs. Difference}
Concatenation-based methods (\dc, \scc) directly concatenate the two posteriors to preserve the full information,
while difference-based methods capture the discrepancy between two versions of posteriors.
We use two approaches to capture this discrepancy: element-wise difference (\dd, \sd) and Euclidean distance (\ed).

Overall, \autoref{fig:feature_constructin_compare} shows that, on one hand, concatenation-based methods perform better on the overfitted model, i.e., ResNet50.
On the other hand, the difference-based methods perform better on the well-generalized model, i.e., logistic regression.
We suspect this is due to the fact that the concatenation-based methods rely on the plain posterior information, which can provide a strong signal for membership inference on the overfitted target model.
This is consistent with the conclusion of previous studies~\cite{SSSS17, SZHBFB19} that \mlleak (which uses plain posterior information) performs well on overfitted target models.
While we can also exploit the difference-based methods to mount the attack on the overfitted target models, the attack signal is not as strong as that of the concatenation-based methods as shown in~\autoref{subfig:feature_rf}.
For the well-generalized target models, exploiting the plain posterior information has shown to perform poorly in terms of membership inference~\cite{SSSS17, SZHBFB19}.
In this case, the discrepancy information between two versions of the posteriors captured by the difference-based methods is more informative than the concatenation-based feature construction methods.

\mypara{Sorted vs.\ Unsorted}
Comparing \dc to \scc and \dd to \sd in \autoref{fig:feature_constructin_compare}, we observe that the attack AUC of both concatenation-based method and difference-based method are clearly better after sorting.
These results confirm our conjecture that sorting could improve the confidence level of the adversary.

\mypara{Feature Selection Summary}
Our empirical comparison provides us with the following rules for the feature construction methods: 
(1) use concatenation-based methods on overfitted models; 
(2) use difference-based methods on well-generalized models; 
(3) sort the posteriors before the concatenation and difference operations.

\begin{table}[!t]
 \centering
 \caption{Attack AUC in different overfitting levels.}
 \footnotesize
 \resizebox{1.0\linewidth}{!}{
 \setlength{\tabcolsep}{0.12em}
 \renewcommand{\arraystretch}{1.3}
 \begin{tabular}{c c | c c | c}
  \toprule
  \textbf{Dataset} & \textbf{\model{o}{}} & \textbf{Train / Test Acc.} & \textbf{Overfitting} & \textbf{AUC / Base-AUC} \\ 
  \toprule
    & LR & 0.795 / 0.782 & 0.013 & 0.600 / 0.505 \\
    \rowcolor{mygray}
    \cellcolor{white} 
    & DT & 0.853 / 0.834 & 0.019 & 0.882 / 0.497 \\
    & RF & 0.852 / 0.843 & 0.009 & 0.659 / 0.459 \\
    \rowcolor{mygray}
    \cellcolor{white} 
    \multirow{-4}{*}{\rotatebox[origin=c]{90}{Adult}} 
    & MLP & 0.767 / 0.763 & 0.004 & 0.506 / 0.503 \\
    \midrule
    & LR & 0.702 /  0.698 & 0.002 & 0.538 / 0.494 \\
    \rowcolor{mygray}
    \cellcolor{white} 
    & DT & 0.722 / 0.701 & 0.021 & 0.929 / 0.501 \\
    & RF & 0.730 / 0.709 & 0.021 & 0.78 / 0.499 \\
    \rowcolor{mygray}
    \cellcolor{white} 
    \multirow{-4}{*}{\rotatebox[origin=c]{90}{Accident}}
    & MLP & 0.670 / 0.644 & 0.026 & 0.513 / 0.493 \\
    \midrule
    & LR  & 0.508 / 0.439 & 0.069 & 0.983 / 0.490 \\
    \rowcolor{mygray}
    \cellcolor{white}
    & DT & 0.404 / 0.373 & 0.031 & 0.941 / 0.503 \\
    & RF & 0.523 / 0.442 & 0.081 & 0.685 / 0.551 \\
    \rowcolor{mygray}
    \cellcolor{white} 
    \multirow{-4}{*}{\rotatebox[origin=c]{90}{Insta-NY}}
    & MLP & 0.738 / 0.483 & 0.255 & 0.619 / 0.553\\
    \midrule
    \multirow{1}{*}{MNIST} 
    & SimCNN & 0.954 / 0.951 & 0.003 & 0.511 / 0.496 \\
    \midrule
    & DenseNet & 0.942 /  0.477 & 0.465 & 0.881 / 0.630 \\
    \rowcolor{mygray}
    \cellcolor{white}
    \multirow{-2}{*}{CIFAR10} 
    & ResNet50 & 0.975 / 0.592 & 0.383 & 0.719 / 0.548 \\
  \bottomrule
 \end{tabular}
 }
  \label{table:overfitting}
\end{table}

\subsection{Impact of Overfitting}
\label{subsec:overfitting}

Overfitting measures the difference of accuracy between training and testing data.
Previous studies~\cite{SSSS17,YGFJ18,LBWBWTGC18} have shown that overfitted models are more susceptible to \mlleak attacks, while well-generalized models are almost immune to them.
In this subsection, we want to revisit the impact of overfitting on \method.

\autoref{table:overfitting} depicts the attack AUC for different overfitting levels.
We use random forest as attack model, and use \sd and \scc as feature construction method for well-generalized and overfitted target model, respectively.
In general, our attack consistently outperforms the classicial membership inference on both well-generalized and overfitted models.
On the overfitted models, i.e., CIFAR10 datasets with ResNet50 and DenseNet as target model, we can observe that the classical membership inference also works. However, our attack can achieve much better performance.
On the other hand, the experimental results show that \textbf{\method can still correctly infer the membership status of the \target sample in well-generalized models}. 
For example, when the target model is a decision tree, the overfitting level in Adult (Income) dataset is $0.019$, thus decision tree can be regarded as a well-generalized model.
While the performance of \mlleak on this model is equivalent to random guessing (AUC = $0.497$),
\method performs very well, with an AUC of $0.882$. 

In summary, our attack performance is relatively independent of the overfitting level.

\begin{figure*}[!t]
\centering
    \begin{minipage}{1\textwidth}
    \begin{subfigure}{0.33\columnwidth}
    \includegraphics[width=\columnwidth]{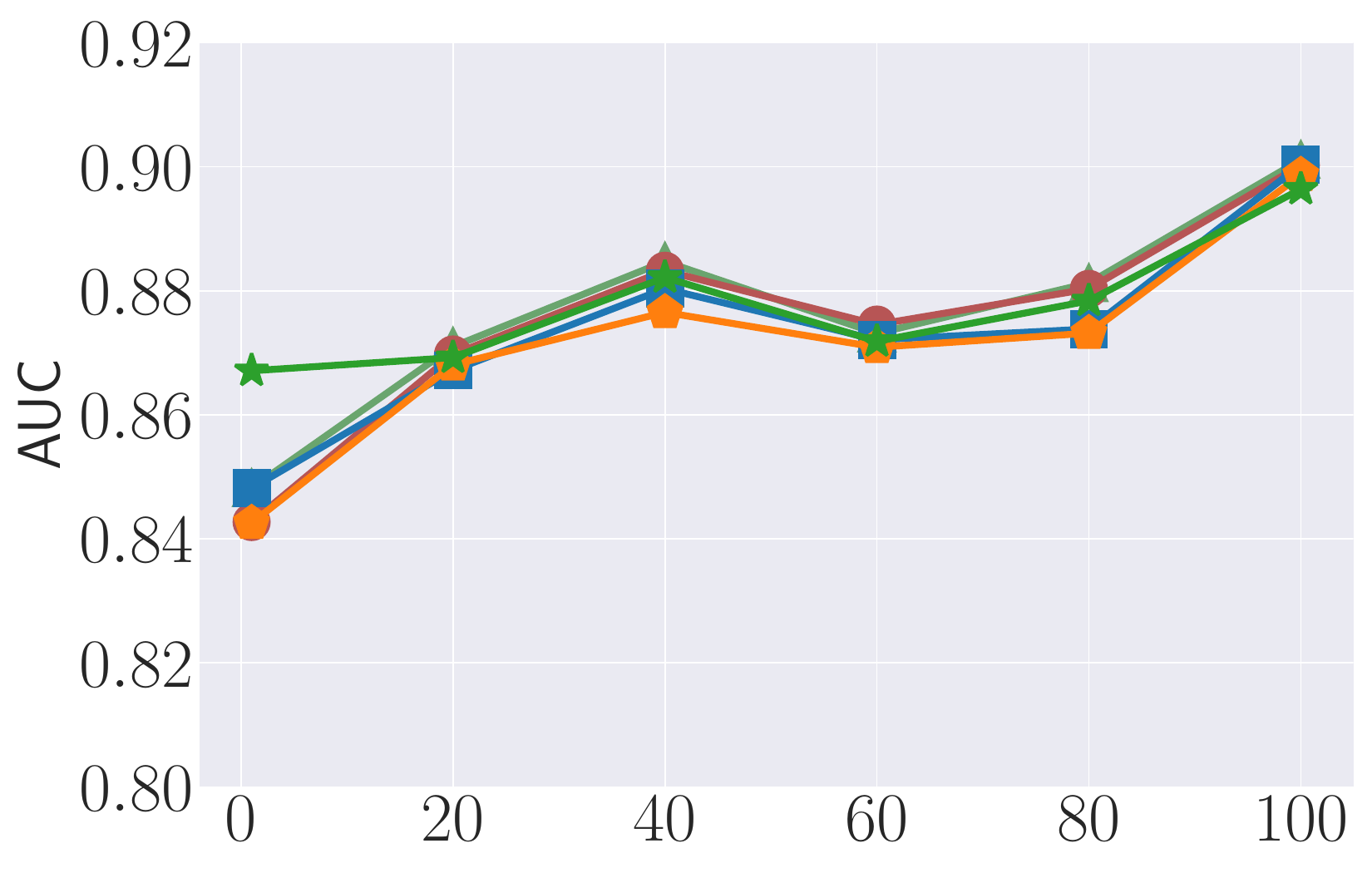}
    \caption{Number of Shadow Original Models \model{o}{}}
    \label{subfig:hyper_sn_DT}
    \end{subfigure}
    \begin{subfigure}{0.33\columnwidth}
    \includegraphics[width=\columnwidth]{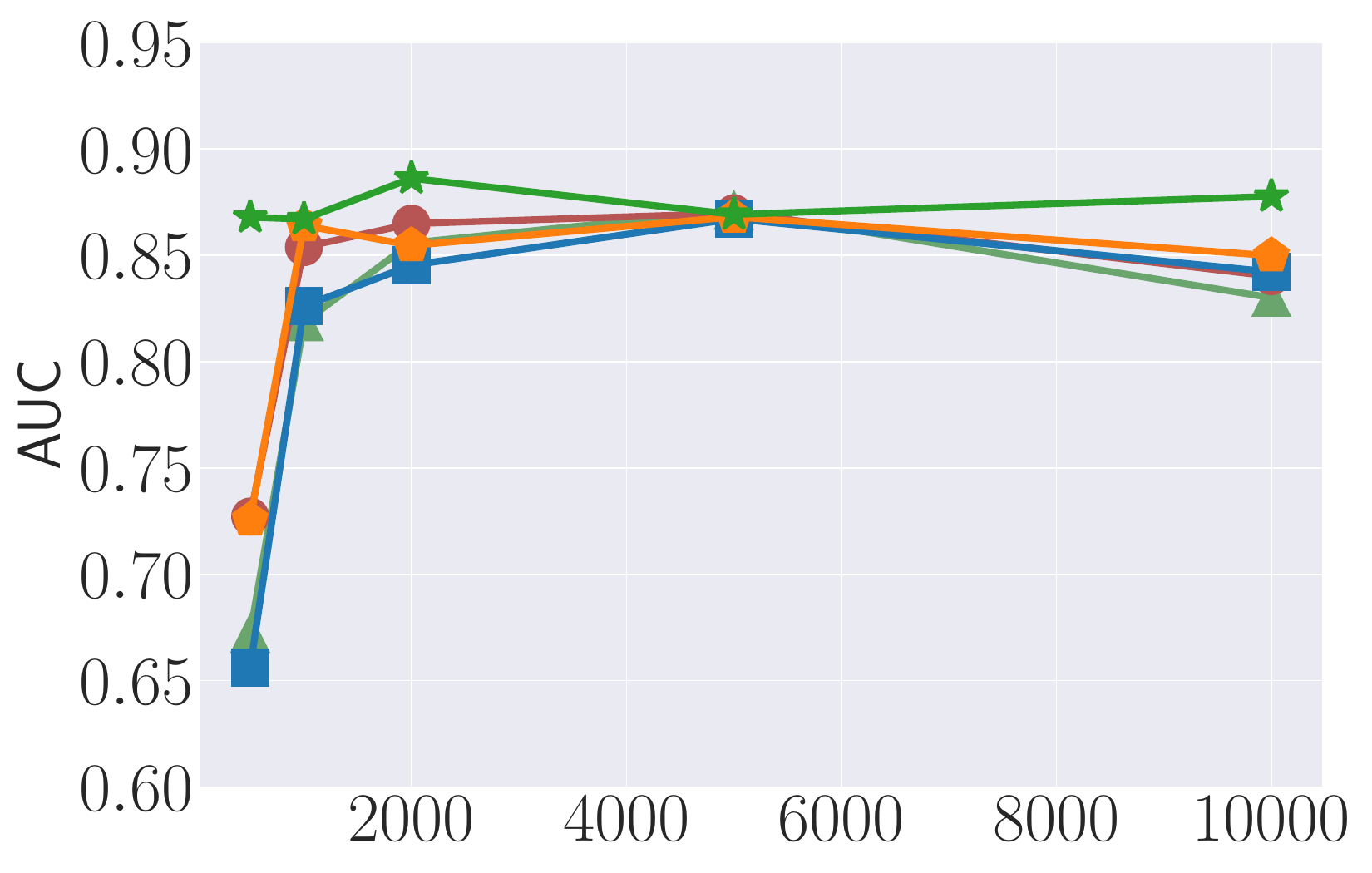}
    \caption{Number of Training Samples}
    \label{subfig:hyper_ss_DT}
    \end{subfigure}
    \begin{subfigure}{0.33\columnwidth}
    \includegraphics[width=\columnwidth]{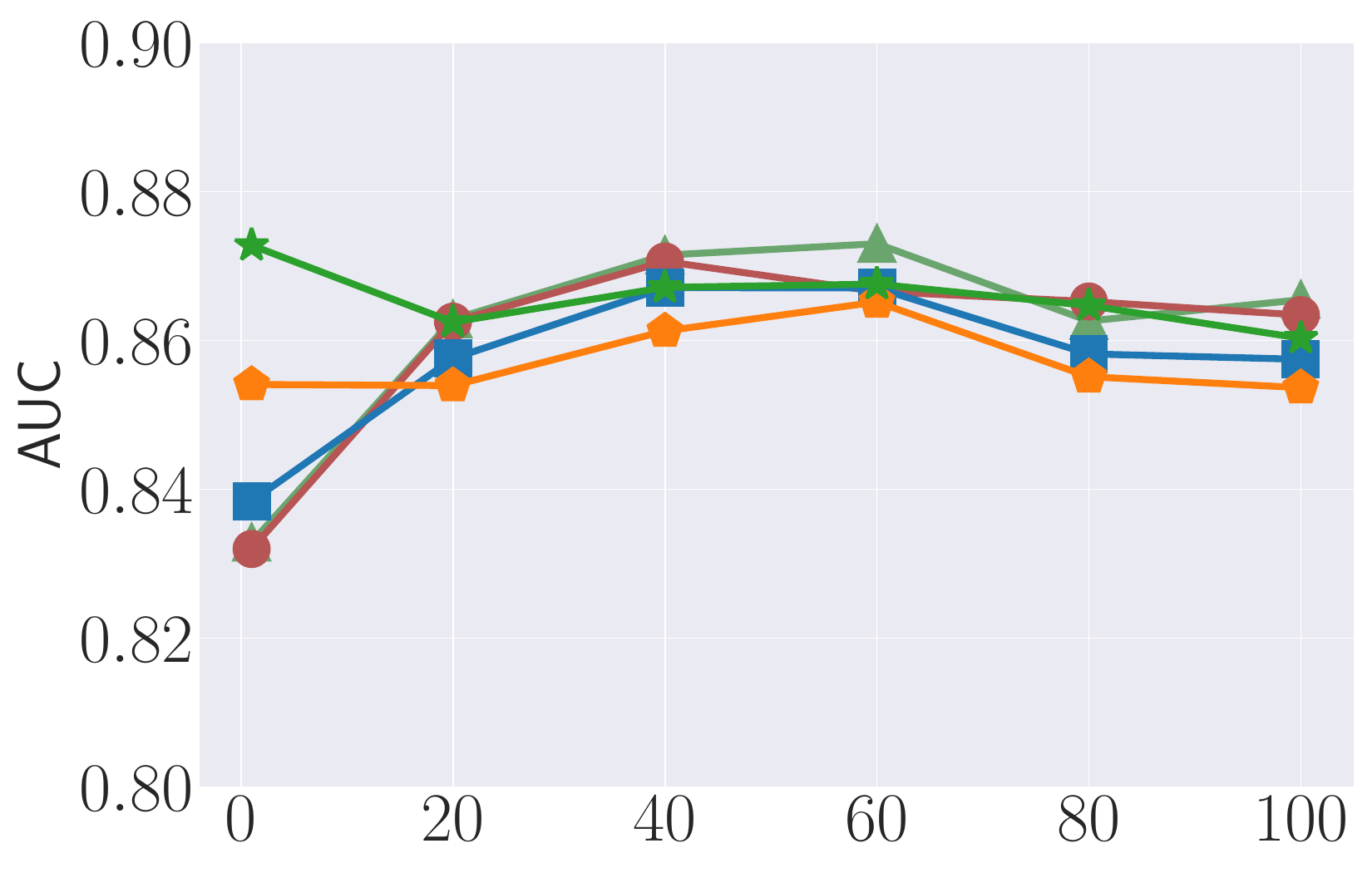}
    \caption{Number of Unlearned Models per \model{o}{s}}
    \label{subfig:hyper_su_DT}
    \end{subfigure}
    \medskip
    \end{minipage}
    \subfloat{\includegraphics[width=0.8\textwidth]{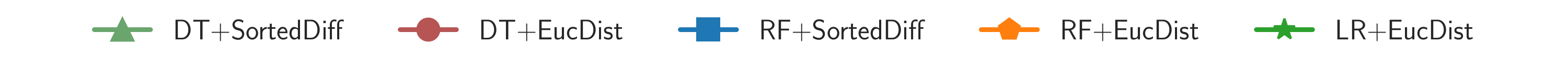}} \\
    [-1ex]
    \caption{Attack AUC sensitivity to different hyperparameters on Adult (income) dataset with decision tree as target model. 
    The legends stand for $5$ combinations of attack models and feature construction methods guided by \autoref{subsec:choice_feature}.
    }
    \label{fig:hyper_income_DT}
\end{figure*}

\begin{table}[!tpb]
 \centering
 \caption{Attack AUC for dataset and model transfer.
 Names in the left of arrows stand for configurations of shadow model.
 Values in the parentheses stand for the attack AUC of the \mlleak.
 Columns stand for dataset transfer, rows stand for model transfer.}
 \label{table:transfer_attack}
 \footnotesize
 \resizebox{1.0\linewidth}{!}{
 \setlength{\tabcolsep}{0.4em}
 \renewcommand{\arraystretch}{1.3}
 \begin{tabular}{c c c}
 \toprule
    Shadow$\rightarrow${Target} & \textbf{Insta-NY$\rightarrow${Insta-NY}} & Insta-NY$\rightarrow${Insta-LA} \\ 
    \toprule
    \rowcolor{mygray}
    \textbf{DT$\rightarrow${DT}} & \textbf{0.944 (0.491)} & 0.931 (0.503)\\
    DT$\rightarrow${LR} & 0.964 (0.494) & 0.974 (0.513) \\
    \rowcolor{mygray}
    \textbf{LR$\rightarrow${LR}} & \textbf{0.986 (0.505)} & 0.982 (0.511) \\
    LR$\rightarrow${DT} & 0.927 (0.502) & 0.926 (0.508) \\
    \bottomrule
    \toprule
    Shadow$\rightarrow${Target} & \textbf{CIFAR10$\rightarrow${CIFAR10}} & CIFAR10$\rightarrow${STL10} \\ 
    \toprule
    \rowcolor{mygray}
    \textbf{DenseNet $\rightarrow${DenseNet}} & \textbf{0.881 (0.630)} & 0.813 (0.621)\\
    DenseNet$\rightarrow${ResNet50} & 0.847 (0.624) & 0.805 (0.632) \\
    \rowcolor{mygray}
    \textbf{ResNet50$\rightarrow${ResNet50}} & \textbf{0.719 (0.548)} & 0.687 (0.550) \\
    ResNet50$\rightarrow${DenseNet} & 0.721 (0.523) & 0.675 (0.542) \\
  \bottomrule
 \end{tabular}
 }
\end{table}

\subsection{Ablation Study}
\label{subsec:hyperparameter}

We now evaluate the impact of the hyperparameters on the performance of \method.
Specifically, we focus on the number of shadow original models \so, the number of samples \sr per shadow original model, and the number of \unlearned models \su per shadow original model.
The corresponding hyperparameters of the target models are fixed (as defined at the end of \autoref{subsec:experiment_setup}), since only the hyperparameters of the shadow models can be tuned to launch the attack.

We conduct the experiments on Adult (Income) dataset with decision tree as target model.
Following our findings in \autoref{subsec:choice_feature}, we evaluate
the attack AUC of different combinations of attack models, i.e., decision tree, random forest and logistic regression, and difference-based feature construction methods, i.e., \dd, \sd, \ed.

\mypara{Number of Shadow Original Models \so}
\autoref{subfig:hyper_sn_DT} depicts the impact of \so, which varies from $1$ to $100$.
The figure shows that the attack AUC sharply increases when \so increases from $1$ to $5$, but remains quite stable for greater values of \so.
This indicates that setting $\so=5$ is enough for the diversity of the shadow original models.

\mypara{Number of Samples \sr per Model}
\autoref{subfig:hyper_ss_DT} illustrates the impact of \sr $\in \{500, 1000, 2000, 5000, 10000\}$.
When \sr increases from $500$ to $1000$, the attack AUC with \sd increases from $0.67$ to $0.83$, while the attack AUC with \ed increases from $0.73$ to $0.86$, {except for logistic regression}. However, adding more than 1000 samples does not help improve the attack performance further.

\mypara{Number of Unlearned Models \su per Shadow Original Model}
\autoref{subfig:hyper_su_DT} illustrates the impact of \su, which varies from $1$ to $100$.
We observe that \su has negligible impact on the attack AUC.
This indicates that using a few unlearned models is sufficient to achieve a high attack performance.

\begin{figure}[!t]
    \centering
    \begin{subfigure}{0.85\columnwidth}
    \includegraphics[width=\textwidth]{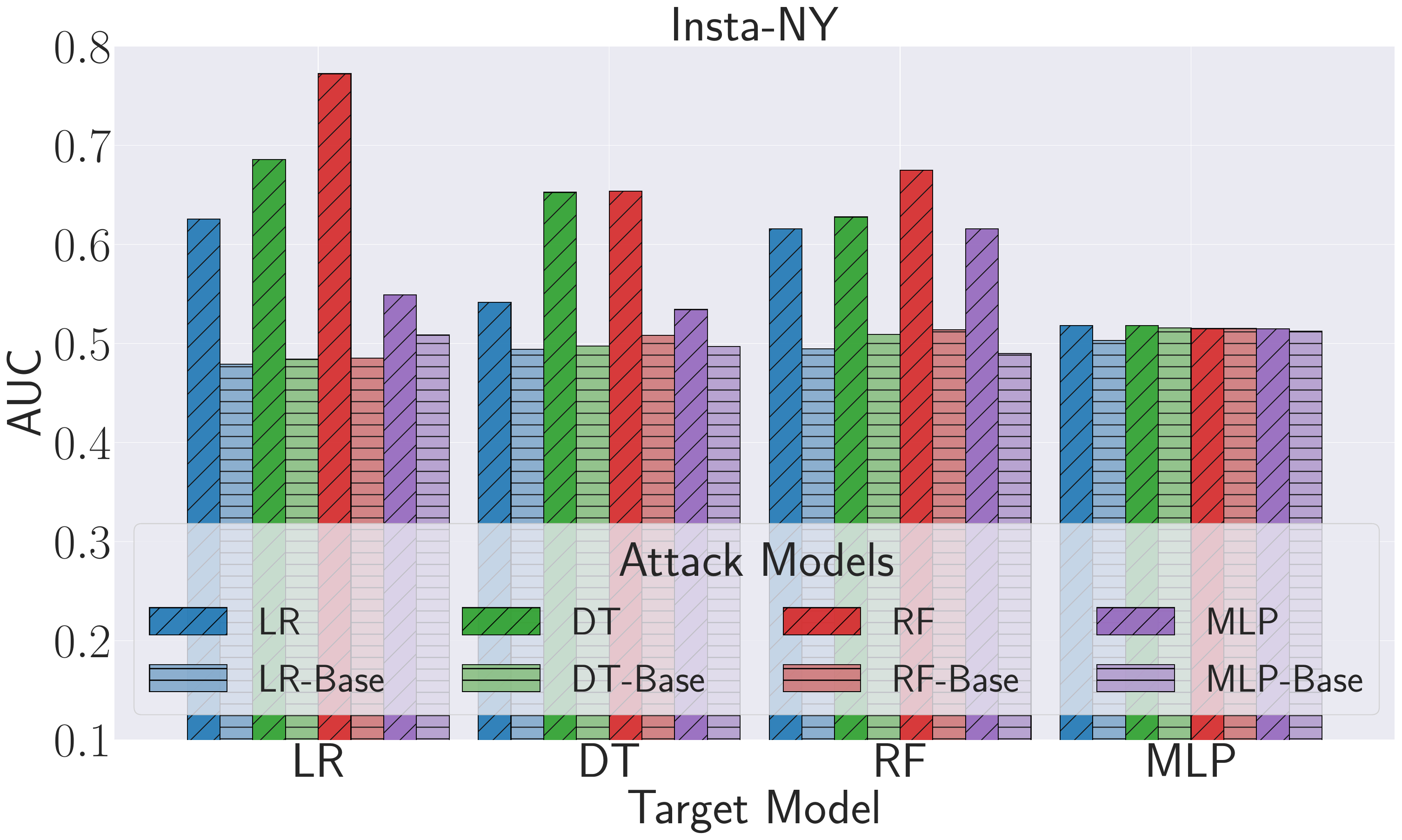}
    \label{subfigure:compare_sisa}
    \end{subfigure}
    \vspace{-0.5cm}
    \caption{Attack AUC for the \sisa method on the Insta-NY dataset.
    The transparent bars stand for the \mlleak.
    }
    \label{fig:compare_sisa}
\end{figure}

\subsection{Attack Transferability}
\label{subsec:transfer_attack}

In practice, the adversary might not able to get the same distribution dataset or same model structure to train the shadow models.
We next validate the dataset and model transferability between shadow model and target model.
That is, we evaluate whether the adversary can use a different dataset and model architecture than the target model to train the shadow models.
We evaluate on both categorical datasets with simple model structure and image dataset with complex model structure.

\mypara{Dataset Transferability}
Comparing the AUC values of the transfer setting with that of the non-transfer setting, i.e., bold rows in column Insta-NY$\rightarrow$Insta-LA and CIFAR10$\rightarrow$STL10, we only observe a small performance drop for all target models.
For instance, when the target model is decision tree, the attack AUC of transfer setting and non-transfer setting are $0.944$ and $0.931$, respectively. The attack AUC only drops by $1\%$.

\mypara{Model Transferability}
For model transferring attack, we evaluate the pairwise transferability among decision tree and logistic regression.
In \autoref{table:transfer_attack}, unbold rows in column Insta-NY$\rightarrow$Insta-NY and CIFAR10$\rightarrow$CIFAR10 illustrate the performance of model transfer.
The experimental results show that model transfer only slightly degrades the attack performance of \method.
For example, when the shadow model and target model are both LR, the attack AUC equals to $0.986$.
When we change the target model to decision tree, the attack AUC is still of $0.927$.

\mypara{Dataset and Model Transferability}
Unbold rows of unbold columns show the attack AUC when we transfer both the dataset and the model simultaneously.
Even in this setting, \method can achieve pretty good performance.
This result shows robust transferability of \method, when the adversary do not have access to same distribution data and same model architectures.

\subsection{Evaluation of the \sisa Method}

The unlearning algorithm we focused on so far is retraining from scratch, which can become computationally prohibitive for large datasets and complex models.
Several \emph{approximate} unlearning algorithms have been proposed to accelerate the training process.
In this subsection, we evaluate the performance of \method against the most general approximate unlearning algorithm, \sisa~\cite{BCCJTZLP21}.

\mypara{Setup}
We remind the readers that the main idea of \sisa is to split the original dataset into $k$ disjoint shards and train $k$ sub-models.
In the inference phase, the model owner aggregates the prediction of each sub-model to produce the global prediction using some aggregation algorithm.
In this experiment, we set $k=5$ and use posterior average as aggregation algorithm.
\autoref{fig:compare_sisa} illustrates the attack AUC on the Insta-NY dataset.
We report the experimental results of four different target models and four different attack models.
For each attack model, we select the best features following the principles described in \autoref{subsec:choice_feature}.

\mypara{Results}
The experimental results show that \method performance drops compared to the \scratch algorithm. 
We posit this is because the aggregation algorithm of \sisa reduces the influence of a specific sample on its global model.
This observation further motivates the deployment of unlearning methods such as \sisa in real-world applications.

\begin{figure*}[!t]
\centering
    \begin{minipage}{1\textwidth}
    \begin{subfigure}{0.33\columnwidth}
    \includegraphics[width=\columnwidth]{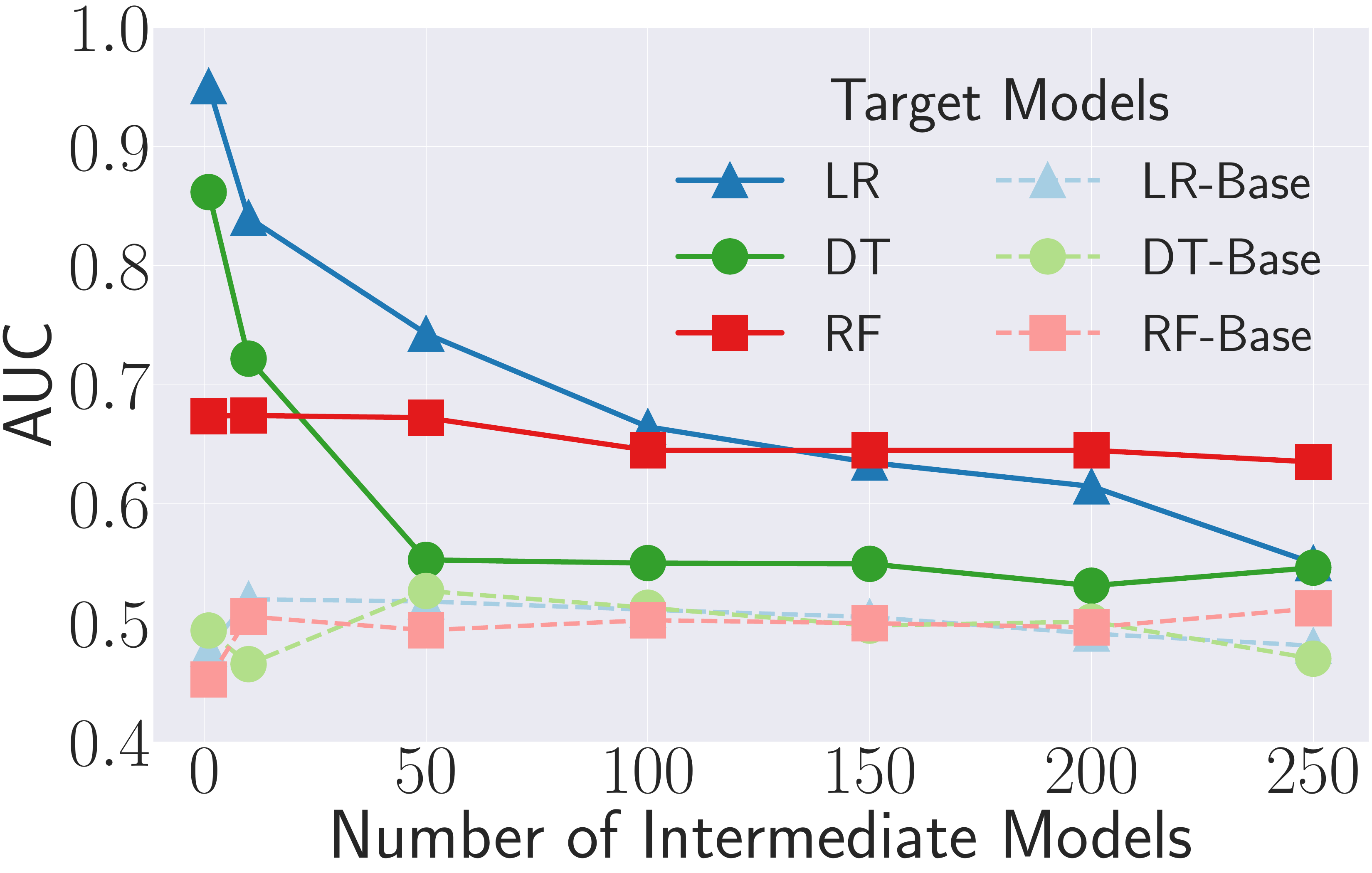}
    \caption{Multiple Intermediate Unlearned Models}
    \label{subfigure:multi_version}
    \end{subfigure}
    \begin{subfigure}{0.33\columnwidth}
    \includegraphics[width=\columnwidth]{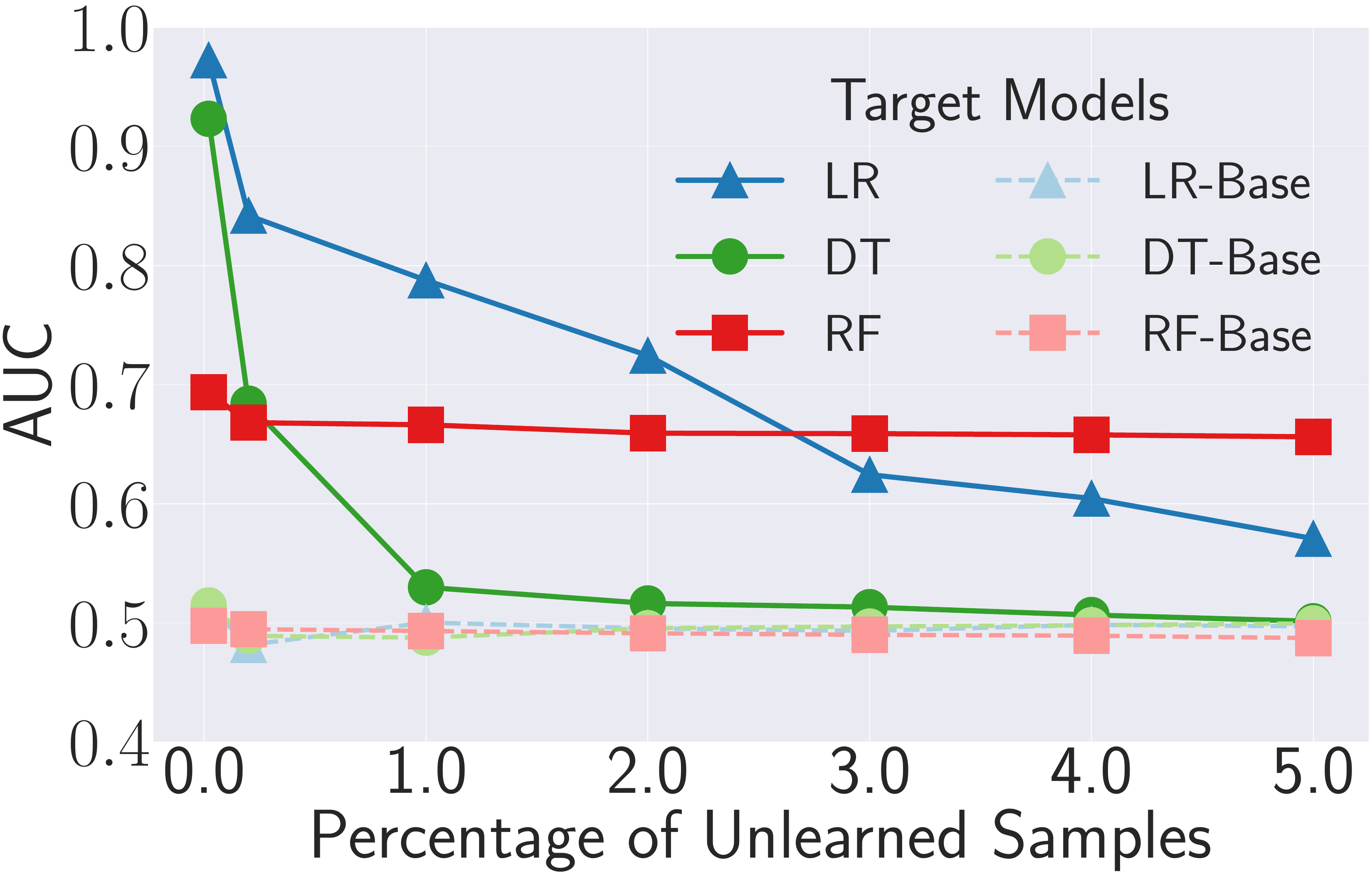}
    \caption{Group Deletion}
    \label{subfigure:multi_group}
    \end{subfigure}
    \begin{subfigure}{0.33\columnwidth}
    \includegraphics[width=\columnwidth]{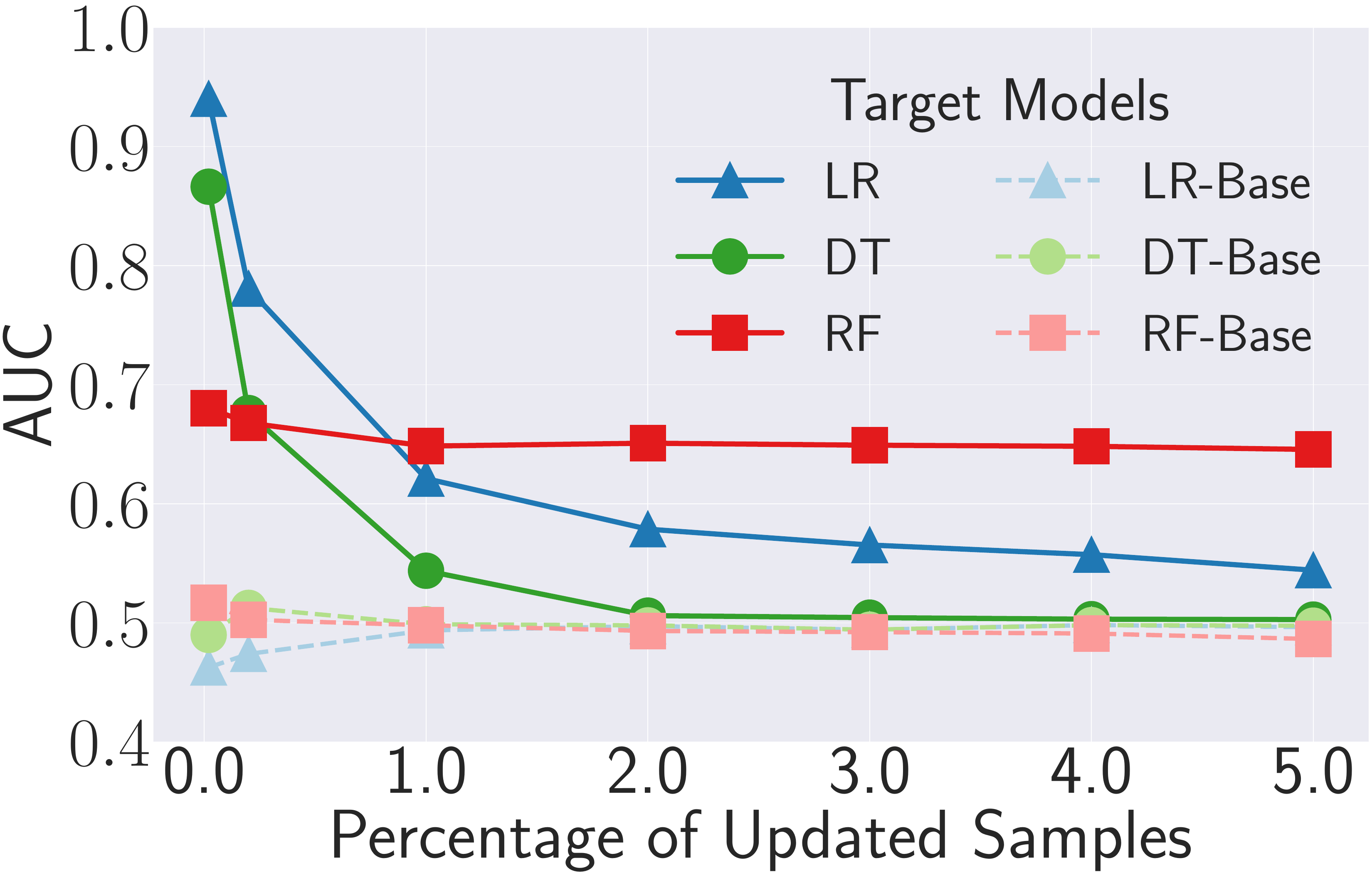}
    \caption{Online Learning}
    \label{subfigure:multi_replace}
    \end{subfigure}
    \medskip
    \end{minipage}
    \\ [-1ex]
    \caption{
    Attack AUC under different scenarios on Insta-NY.
    The dashed lines stand for the attack AUC of classical membership inference.
    Due to space limitation, we omit the results for other models and datasets that hold similar conclusions.
    }
    \label{fig:realworld_scenarios}
\end{figure*}

\section{Attack Under Different Scenarios}
\label{sec:practical_deployment}

Next, we evaluate the effectiveness of our attack in different scenarios that might exist in practice.
We first focus on the case when there exists multiple intermediate versions of unlearned models.
Second, we consider when a group of samples are deleted.
Third, we investigate the online learning setting when multiple samples are deleted and added simultaneously.
Finally, we evaluate the impact of unlearning on remaining samples' membership privacy.
The experimental setup is the same as in~\autoref{sec:exp}.

\subsection{Multiple Intermediate Unlearned Models}
\label{subsec:multi_version_deletion}

\mypara{Intermediate Models}
As discussed in the threat model (\autoref{subsec:threat_model}), the adversary gains access to the original model and unlearned model by continuously querying the black-box target model.
In practice, the adversary obtains access to two consecutive versions of models by two consecutive queries. 
However, the model owner would produce an unlearned model every time when it receives deletion requests; thus, there might be multiple unlearned models between these two consecutive queries that is unknown to the adversary. 
We call these models \textit{intermediate models}. 
Here, we evaluate the effectiveness of our attack when multiple intermediate unlearned models exist.

\mypara{Setup}
Due to space limitation, we concentrate on the Insta-NY dataset with three different target models, while the conclusions are consistent for other datasets.
We use LR as the attack model and select the best features following the principles described in \autoref{subsec:choice_feature}.
\autoref{subfigure:multi_version} depicts the results.
The x-axis represents the number of intermediate unlearned models we studied, i.e., $\{1, 10, 50, 100, 250\}$.

\mypara{Results}
The experimental results show that our attack consistently degrades privacy of the target sample comparing with the classical membership inference.
In addition, the attack AUC drops when the number of intermediate models increases.
This is expected since the impact of the target sample is masked by previously deleted samples.
That is, if there exist multiple intermediate models, the discrepancy information between the original model and the unlearned model is contributed by both the target sample and other deleted samples corresponding to the intermediate models.
In other words, the impact of the target sample and other deleted samples is entangled with each other, making the inference of the membership status of the target sample more difficult.

Note that the data samples are unlikely to be revoked very frequently in practice, and the number of intermediate models are unlikely to be very large, which means our attack is still effective in real-world settings.
For instance, our attack AUC can achieve at least $0.84$ when the number of intermediate models are less than $10$ when the target model is LR.
 
\subsection{Group Deletion}
\label{subsec:group_deletion}

In practice, there could exist cases where a group of samples are deleted at once before generating the \unlearned model.
This can happen when multiple data owners request the deletion at the same time, or when the model owner caches the deletion requests and updates the model only when he has received numerous requests to save computational resources.

\mypara{Setup}
We conduct experiments on \method in the group deletion scenario.
We randomly delete a group of data samples from each original model to generate the \unlearned model.
The ratio of samples in each group takes value from $\{0.02\%, 0.2\%, 1\%, 2\%, 5\%\}$.
We delete at most $5\%$ of the data samples since in practice it is unlikely that more than $5\%$ of users revoke their data.
We evaluate our attack on the Insta-NY dataset with three target models.
Notice that the unlearned model of the group deletion is the same as in~\autoref{subsec:multi_version_deletion} when the group size equals the number of intermediate unlearned models.
The difference is that in group deletion, we consider all samples in the group as target samples.

\mypara{Results}
\autoref{subfigure:multi_group} shows that our consistently outperforms classical membership inference attack, demonstrating extra information leakage in group deletion.
However, the attack performance of group deletion is slightly worse than single sample deletion, even though our attack is still effective when the group size is smaller than $0.2\%$.
For example, when the target model is LR, the attack AUC of single deletion and group deletion ($0.2\%$ target samples) are $0.972$ and $0.842$, respectively.
The reason is that a single sample could be hidden among the group of deleted samples, thereby preserving its membership information.
This result reveals that conducting group deletion could mitigate, to some extent, the impact of \method.

In practice, we believe 0.2\% might already be too large for unlearning. 
The results of~\cite{BBCCCFGHHIDKLNNOPSTV19} show that 3.2 million requests for removing URLs have been issued to Google for 5 years which certainly constitutes less than 0.2\% of the total URLs Google indexes.

\subsection{Online Learning}
\label{subsec:online_learning}

In real-world deployments, ML models are often updated with new samples, which is known as online learning or incremental learning.
Next, we evaluate the performance of our attack in online learning settings where multiple samples are deleted and added simultaneously.

\mypara{Setup}
To set up the experiment, we delete a group of target samples from the original dataset and add the same number of new samples; then we retrain the model from scratch to obtain the unlearned model.
We conduct experiments on Insta-NY with different target models and use LR as the attack model.

\mypara{Results}
\autoref{subfigure:multi_replace} show that adding samples to the target model in the unlearning process has slight impact on our attack.
For example, compared with purely deletion, the attack AUC only slightly drops from $0.972$ to $0.940$ when the target model is LR and the number of unlearned samples equals to $10$ ($0.2\%$).

\subsection{Impact on Remaining Samples}
\label{subsec:non_redacted_samples}

In the end, we evaluate whether deleting the target sample can influence the privacy of other remaining samples.

\mypara{Setup}
We use the same attack pipeline described in \autoref{subsec:pipeline} to mount the attack.
Concretely, we use data samples that reside in both the original model and unlearned model as positive cases, and use shadow/target negative dataset as negative cases.
We concentrate on the Insta-NY dataset with four target models and four attack models where one data sample is deleted.

\begin{figure}[!t]
    \centering
    \begin{subfigure}{0.85\columnwidth}
    \includegraphics[width=\textwidth]{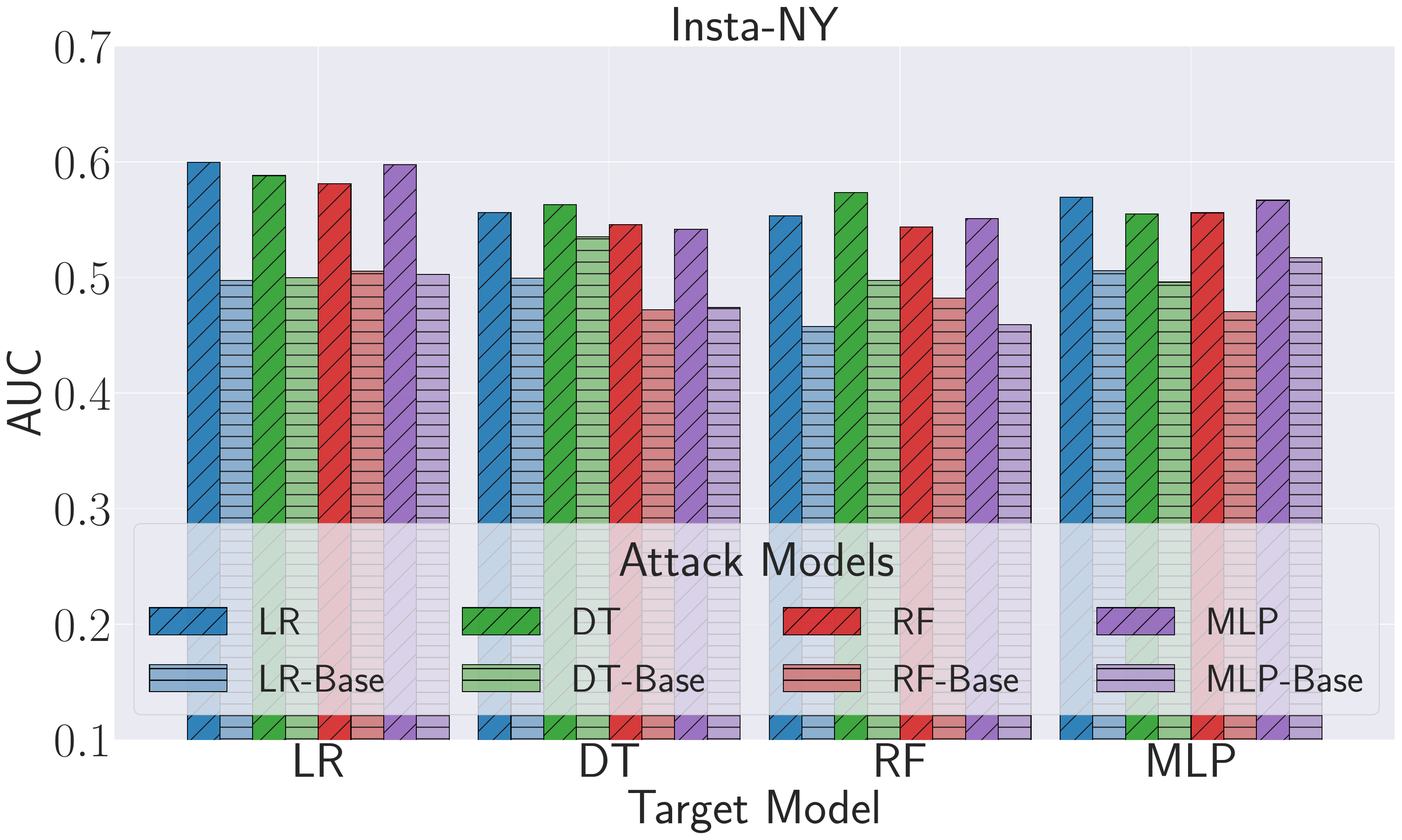}
    \label{subfigure:in_in}
    \end{subfigure}
    \vspace{-0.7cm}
    \caption{Attack AUC of the remaining samples on the Insta-NY dataset.
    The transparent bars stand for the \mlleak.}
    \label{fig:in_in}
\end{figure}

\mypara{Results}
\autoref{fig:in_in} shows that the attack AUC of our attack is higher than that of the classical membership inference, which only exploit information of the original model, indicating deleting the target sample also degrades privacy, to some extent, of the remaining samples.
However, the attack AUC of all target models are less than $0.6$, meaning the remaining samples are less sensitive to our attack.
This is expected due to the fact that the remaining samples are members of both the original model and unlearned model. 
Deleting other data samples has some but limited impact on their posteriors in the unlearned model.

\subsection{Takeaways}

Through our extensive experiments in \autoref{sec:exp} and \autoref{sec:practical_deployment}, we have made the following important observations:

\begin{itemize}
    \item Our attack consistently degrades the membership privacy of the \scratch unlearning method compared to classical membership inference.
    The attack performance drops for the \sisa unlearning method, which motivates the deployment of unlearning methods such as \sisa in real-world applications.
    \item We obtain the following rules for selecting the feature construction methods: 
    (1) use concatenation-based methods on overfitted models; 
    (2) use difference-based methods on well-generalized models; 
    (3) sort the posteriors before the concatenation and difference operations.
    \item Our transferring attacks show that our attack is still effective when the shadow model is trained on different-distributed datasets and different architecture from the target model.
    \item When the number of unlearned/updating samples represents less than 0.2\% of the training dataset, our attack is still effective in the scenarios of multiple intermediate unlearned models, group deletion, and online learning.
    \item Deleting the target sample also degrades privacy of the remaining samples to some extent; however, the remaining samples are less sensitive to our attack.
\end{itemize}

\section{Possible Defenses}
\label{sec:defense}

In this section, we explore four possible defense mechanisms and empirically evaluate their effectiveness.
The former two mechanisms reduce the information accessible to the adversary~\cite{SSSS17}, and the latter two eliminate the impact of a single sample on the output of the ML models.

\begin{table}[!t]
    \centering
    \caption{Attack AUC of the defense mechanisms.
    We list the attack performance of no defense mechanisms (ND), publishing the Top-k confidence values (Top-1, Top-2, Top-3), label-only defense (Label), temperature scaling (TS) based defense, and differential privacy (DP) based defense.
    \dset{o}{}, \model{T}{}, and \model{A}{} stand for original dataset, target model and attack model, respectively.
    For DP, we set $\delta=10^{-5}, \epsilon_1=4.64, \epsilon_2=0.7$.}
    \resizebox{1.0\linewidth}{!}{
    \setlength{\tabcolsep}{0.12em}
    \renewcommand{\arraystretch}{1.3}
    \begin{tabular}{c  c | c c c c c c c c}
    \toprule
    \textbf{\dset{o}{} (\model{T}{})} & \textbf{\model{A}{}} & \textbf{ND} & \textbf{Top-1}  & \textbf{Top-2} & \textbf{Top-3} & \textbf{Label} & \textbf{TS} & \textbf{DP[$\epsilon_1$]} & \textbf{DP[$\epsilon_2$]} \\
    \toprule
    \rowcolor{mygray}\cellcolor{white}
    & RF & 0.916 & 0.899 & 0.906 & 0.911 & 0.501 & - & - & - \\
    & DT & 0.918 & 0.903 & 0.906 & 0.910 & 0.506 & - & - & - \\
    \rowcolor{mygray}\cellcolor{white}
    & LR & 0.918 & 0.904 & 0.907 & 0.911 & 0.506 & - & - & - \\
    \multirow{-4}{*}{\rotatebox[origin=c]{90}{Adult (DT)}}
    & MLP & 0.918 & 0.904 & 0.909 & 0.907 & 0.493 & - & - & - \\
    \midrule
    \rowcolor{mygray}\cellcolor{white}
    & RF & 0.937 & 0.930 & 0.931 & 0.942 & 0.506 & - & - & - \\
    & DT & 0.938 & 0.932 & 0.932 & 0.943 & 0.502 & - & - & - \\
    \rowcolor{mygray}\cellcolor{white}
    & LR & 0.928 & 0.923 & 0.927 & 0.926 & 0.502 & - & - & - \\
    \multirow{-4}{*}{\rotatebox[origin=c]{90}{Insta-NY (DT)}}
    & MLP & 0.928 & 0.923 & 0.927 & 0.929 & 0.505 & - & - & - \\
    \midrule
    \rowcolor{mygray}\cellcolor{white}
    & RF & 0.976 & 0.947 & 0.965 & 0.965 & 0.546 & 0.635 & 0.519 & 0.477 \\
    & DT & 0.972 & 0.946 & 0.961 & 0.961 & 0.546 & 0.654 & 0.524 & 0.500 \\
    \rowcolor{mygray}\cellcolor{white}
    & LR & 0.969 & 0.948 & 0.960 & 0.962 & 0.546 & 0.610 & 0.519 & 0.500 \\
    \multirow{-4}{*}{\rotatebox[origin=c]{90}{Insta-NY (LR)}}
    & MLP & 0.970 & 0.948 & 0.960 & 0.966 & 0.453 & 0.653 & 0.506 & 0.504 \\
    \bottomrule
    \end{tabular}
    }    
    \label{table:defense}
\end{table}

\mypara{Publishing the Top-$k$ Confidence Values}
This defense reduces the attacker's knowledge by only publishing top $k$ confidence values of the posteriors returned by both original and \unlearned models.
Formally, we denote the posterior vector as $\post{}{} = [p_1, p_2, \cdots, p_\ell]$, where $\ell$ is the number of classes of the target model and $p_i$ is the confidence value of class $i$.
When the target model receives a query, the model owner calculates posteriors \post{}{} and sorts them in descending order, resulting in $\post{}{s} = [p_1^s, p_2^s, \cdots, p_\ell^s]$.
The model owner then publishes the first $k$ values in \post{}{s}, i.e., $[p_1^s, p_2^s, \cdots, p_k^s]$.

In the machine unlearning setting, the top $k$ confidence values of the original model and the \unlearned model may not correspond to the same set of classes.
To launch \method, the adversary constructs a \textit{pseudo-complete posterior vector} for both original model and \unlearned model.
The pseudo-complete posteriors take the published confidence values for their corresponding classes, and evenly distributes the remaining confidence value to other classes, i.e., for $j\in\{k+1,\ldots,\ell\}$, $p^s_j = \frac{1 - (p_1^s + p_2^s + \cdots + p_k^s)}{\ell - k}$.
The adversary can then launch \method using the pseudo-complete posteriors.

\autoref{table:defense} shows the experimental results of Top-$1$, Top-$2$ and Top-$3$ defenses on Insta-NY and Adult.
For the Adult dataset, we report the results of decision tree as the target model; for the Insta-NY dataset, we report the results of decision tree and logistic regression as the target model.
We report the performance of $4$ different attack models, each selecting the best feature following the principle described in \autoref{subsec:choice_feature}.
The results show that publishing top $k$ confidence value \emph{cannot} effectively mitigate \method.

\mypara{Publishing the Label Only}
This defense further reduces the information accessible to the adversary by only publishing the predicted label instead of confidence values (posteriors).
To launch \method, the adversary also needs to construct the pseudo-complete posteriors for both the original model and \unlearned model.
The main idea is to set the confidence value of the predicted class as $1$, and set the confidence value of other classes as $0$.
\autoref{table:defense} illustrates the performance of the ``label only'' defense.
The experimental setting is similar to Top-$k$ defense.
The experimental results show that the ``label only'' defense can effectively mitigate \method in all cases.
The reason is that deleting one sample is unlikely to change the output label of a specific \target sample.

It is worth noting that recent studies have shown that an adversary can recover to a large extent the posteriors from the label with the so-called sampling attack~\cite{ROF20,LZ21}.
In this case, our membership inference attack is still effective. 
We leave the investigation of the improved attack in the presence of label-only publishing defense as future work.

\mypara{Temperature Scaling} 
Temperature scaling divides the logits vector by a learned scaling parameter, which is a simple yet effective approach to eliminate the over-confident problem of the output posteriors of neural networks~\cite{GPSW17}.
This defense reduces the impact of a single sample on the output posteriors.

\autoref{table:defense} illustrates the performance of the ``temperature scaling'' defense.
We report the performance of $4$ different attack models, each selecting the best feature following the principle described in \autoref{subsec:choice_feature}.
The experimental results show that temperature scaling is an effective defense mechanism.
However, this method is only applicable to neural networks whose last layer is softmax.
Logistic regression in our experiment can be regarded as a neural network with one input layer and one softmax layer.

\mypara{Differential Privacy (DP)}
DP~\cite{LLSY16,DR14,PSMRTE18,BDKU20,NSTPC21} guarantees that any single data sample in a dataset has limited impact on the output.
Previous studies have shown DP can effectively prevent classical membership inference attacks~\cite{JE19,LWHSZBCFZ21}.
To validate whether DP can prevent our membership inference attack in the machine unlearning setting, we train both the original model and unlearned model in a differentially private manner.

We experiment with Differentially-Private Stochastic Gradient Descent (DP-SGD)~\cite{ACGMMTZ16}, the most representative DP mechanism for protecting machine learning models.
The core idea of DP-SGD is to add Gaussian noise to the gradient $g$ during the model training process, i.e., $\Tilde{g} = g + \calN\left(0, \Delta_f^2 \sigma^2 \mathbf{I} \right)$.
We use the Opacus library\footnote{\url{https://github.com/pytorch/opacus}} developed by Facebook to conduct our experiments.
Note that, since DP-SGD can only be applied to the ML models that encounter gradient updating in the training process, we only report the results for logistic regression.
The last two columns of \autoref{table:defense} illustrate the effectiveness of the DP defense.
The experimental results show that DP can effectively prevent our membership inference attack.
It worth noting that DP can inevitably degrade the target model's accuracy. We need carefully tune the privacy budget parameters to strike a trade-off between privacy and model utility in practice.

\medskip
We leave the in-depth exploration of more effective defense mechanisms against \method as a future work.

\section{Related Work}
\label{sec:related}

\mypara{Machine Unlearning}
The notion of machine unlearning is first proposed in~\cite{CY15}, which is the application of the right to be forgotten in the machine learning context.
The most legitimate approach to implement machine unlearning is to remove the revoked samples from the original training dataset and retrain the ML model from scratch.
However, retraining from scratch incurs very high computational overhead when the dataset is large and when the revoke requests happen frequently.
Thus, most of the previous studies in machine unlearning focus on reducing the computational overhead of the unlearning process~\cite{CY15, BCCJTZLP21, BSZ20, ISCZ21}.

For instance, Cao et al. proposed to transform the learning algorithms into summation form that follows statistical query learning, breaking down the dependencies of training data~\cite{CY15}.
To remove a data sample, the model owner only needs to remove the transformations of this data sample from the summations that depend on this sample.
However, this algorithm is not applicable to learning algorithms that cannot be transformed into summation form, such as neural networks.
Bourtoule et al. ~\cite{BCCJTZLP21} proposed a more general algorithm named \sisa.
The main idea of \sisa is to split the training data into disjoint shards, with each shard training one sub-model.
To remove a specific sample, the model owner only needs to retrain the sub-model that contains this sample.
To further speed up the unlearning process, the authors proposed to split each shard into several slices and store the intermediate model parameters when the model is updated by each slice.

Another line of machine unlearning study aims to verify whether the model owner complies with the data deletion request.
Sommer et al.~\cite{SSWM20} proposed a backdoor-based method.
The main idea is to allow the data owners to implant a backdoor in their data before training the ML model in the MLaaS setting.
When the data owners later request to delete their data, they can verify whether their data have been deleted by checking the backdoor success rate. 

The research problem in this paper is orthogonal to previous studies.
Our goal is to quantify the unintended privacy risks for the deleted samples in machine learning systems when the adversary has access to both original model and \unlearned model.
To the best of our knowledge, this paper is the first to investigate this problem.
Although quantifying privacy risks of machine unlearning has not been investigated yet, there are multiple studies on quantifying the privacy risks in the general right to be forgotten setting.
For example, Xue et al.~\cite{XMCAR16} demonstrate that in search engine applications, the right to be forgotten can enable an adversary to discover deleted URLs when there are inconsistent regulation standards in different regions.
Ellers et al.~\cite{ECSSL19} demonstrate that, in network embeddings, the right to be forgotten enables an adversary to recover the deleted nodes by leveraging the difference between the two versions of the network embeddings.

\mypara{Membership Inference}
Shokri et al.~\cite{SSSS17} presented the first membership inference attack against ML models. 
The main idea is to use shadow models to mimic the target model’s behavior to generate training data for the attack model.
Salem et al.~\cite{SZHBFB19} gradually removed the assumptions of~\cite{SSSS17} by proposing three different attack methods.
Since then, membership inference has been extensively investigated in various ML models and tasks, such as federated learning~\cite{MSCS19}, white-box classification~\cite{NSH19}, generative adversarial networks~\cite{HMDC19,CYZF20}, natural language processing~\cite{SS19}, and computer vision segmentation~\cite{HRSF20}.
Another line of study focused on investigating the impact of overfitting~\cite{YGFJ18, LF20} and of the number of classes of the target model~\cite{SSZ20} on the attack performance.

To mitigate the threat of membership inference, a plethora of defense mechanisms have been proposed.
These defenses can be classified into three classes: reducing overfitting, perturbing posteriors, and adversarial training.
There are several ways to reduce overfitting in the ML field, such as $\ell_2$-regularization~\cite{SSSS17}, dropout~\cite{SZHBFB19}, and model stacking~\cite{SZHBFB19}.
In~\cite{LLR21}, the authors proposed to explicitly reduce the overfitting by adding to the training loss function a regularization term, which is defined as the difference between the output distributions of the training set and the validation set.
Jia et al.~\cite{JSBZG19} proposed a posterior perturbation method inspired by adversarial example.
Nasr et al.~\cite{NSH18} proposed an adversarial training defense to train a secure target classifier.
During the training of the target model, a defender’s attack model is trained simultaneously to launch the membership inference attack. 
The optimization objective of the target model is to reduce the prediction loss while minimizing the membership inference attack accuracy.

\mypara{Attacks against Machine Learning}
Besides membership inference attacks, there exist numerous other types of attacks against ML models~\cite{PMJFCS16, TZJRR16,PMGJCS17,TKPGBM17,PMSW18,OASF18,WG18,GWYGB18,GMXSX18,SHNSSDG18,JZJLW18,LMALZWZ19,WYSLVZZ19,LJZWWLW19,LF20,QMR19,SCSRJ20,JCBKP20,SBBFZ20}.
Ganju et al.~\cite{GWYGB18} proposed a property inference attack aiming at inferring general properties of the training data (such as the proportion of each class in the training data).
Model inversion attack~\cite{FLJLPR14, FJR15} focuses on inferring the missing attributes of the target ML model.
A major attack type in this space is adversarial examples~\cite{PMJFCS16,PMGJCS17,TKPGBM17,PMSW18}.
In this setting, an adversary adds carefully crafted noise to samples aiming at misleading the target classifier.
A similar type of attack is the backdoor attack, where the adversary as a model trainer embeds a trigger into the model for them to exploit when the model is deployed~\cite{GDG17,LMALZWZ19,WYSLVZZ19}.
Another line of work is model stealing, Tram{\`e}r et al.~\cite{TZJRR16} proposed the first attack on inferring a model's parameters.
Other works focus on inferring a model's hyperparameters~\cite{OASF18,WG18}.
An interesting future work will be evaluating these attacks under machine unlearning.

\section{Conclusion}
\label{sec:conclusion}
This paper takes the first step to investigate the unintended information leakage in machine unlearning through the lens of membership inference.
We propose several feature construction methods to summarize the discrepancy between the posteriors returned by original model and \unlearned model.
Extensive experiments on five real-world datasets show that our attack in multiple cases outperform the classical membership inference attack on the target sample, especially on well-generalized models.
We further show that we can effectively infer membership information in other scenarios might exist in practice, including the scenario where there are multiple intermediate unlearned models, the scenario where a group of samples (instead of a single one) are deleted together from the original target model, and the online learning scenario where there are samples to be deleted and added simultaneously.
Finally, we present four defense mechanisms to mitigate the newly discovered privacy risks.
We hope that these results will help improve privacy in practical implementation of machine unlearning.

\section*{Acknowledgement}
\label{sec:ack}
We thank our shepherd Gautam Kamath and the anonymous reviewers for their constructive comments.
This work is partially funded by the Helmholtz Association within the project ``Trustworthy Federated Data Analytics'' (TFDA) (funding number ZT-I-OO1 4).
Tianhao Wang is funded by National Science Foundation (NSF) under Grand No.1931443, a Bilsland Dissertation Fellowship, and a Packard Fellowship.

\bibliographystyle{plain}
\bibliography{normal_generated_py3}

\appendix

\section{Datasets}
\label{app:dataset}

\begin{itemize}
    \item \mypara{UCI Adult\footnote{\url{https://archive.ics.uci.edu/ml/datasets/adult.}}}
    This is a widely used categorical dataset for classification.
    It is a census dataset that contains around $50,000$ samples with $14$ features.
    The classification task is to predict whether the income of a person is over $\$50k$, which is a binary classification task.
	\item \mypara{US Accident\footnote{\url{https://www.kaggle.com/sobhanmoosavi/us-accidents}}}
	This is a countrywide traffic accident dataset, which covers $49$ states of the United States.
	This dataset contains around 3M samples.
	We filter out attributes with too many missing values and obtain $30$ valid features.
	The valid features include temperature, humidity, pressure, etc.
	The classification task is to predict the accident severity level which contains $3$ classes.
	\item \mypara{Insta-NY~\cite{BHPZ17}}
	This dataset contains a collection of Instagram users' location check-in data in New York.
	Each check-in contains a location and a timestamp; and each location belongs to a category.
	We use the number of check-ins that happened at each location in each hour on a weekly basis as the location feature vector.
	The classification task is to predict each location's category among $9$ different categories.
	After filtering out locations with less than 50 check-ins, we get 19,215 locations for Insta-NY dataset.
	Later in the section, we also  make use of check-ins in Los Angeles, namely Insta-LA~\cite{BHPZ17}, for evaluating the data transferring attack. This dataset includes 16,472 locations.
	\item \mypara{MNIST\footnote{\url{http://yann.lecun.com/exdb/mnist/}}}
	MNIST is an image dataset widely use for classification.
	It is a 10-class handwritten digits dataset which contains 42,000 samples, each being formatted into a $28\times28$-pixel image.
	\item \mypara{CIFAR10\footnote{\url{https://www.cs.toronto.edu/~kriz/cifar.html}}}
	CIFAR10 is the benchmark dataset used to evaluate image recognition algorithms.
	This dataset contains 60,000 colored images of size $32\times32$, 
	which are equally distributed on the following 10 classes: airplane, automobile, bird, cat, deer, dog, frog, horse, ship, and truck.
	There are 50,000 training images and 10,000 testing images.
	\item \mypara{STL10~\cite{CNL11}}
	STL10 is a 10-class image dataset with each class containing 1,300 images. 
	Classes include airplane, bird, car, cat, deer, dog, horse, monkey, ship, and truck.
\end{itemize}
 
\section{Hyperparameter Settings of Simple Models}
\label{app:model_setting}

We use multiple ML models in our experiments.
All models are implemented by sklearn version 0.22 except for the logistic regression classifier.
For reproduction purpose, we list their hyperparameter settings as follows:

\begin{itemize}
    \setlength\itemsep{-0.25em}
    \item 
    \mypara{Logistic Regression} 
    We implement a single linear logistic regression classifier with PyTorch. Training with Adam optimizer for 100 epochs.
    \item 
    \mypara{Decision Tree} 
    We use Gini index as criterion, set parameter max\_leaf\_nodes as 10, and set other hyperparameters as default.
    \item 
    \mypara{Random Forest} 
    We use Gini index as criterion, use $100$ estimators, set min\_samples\_leaf=30, and set other hyperparameters as default.
    \item 
    \mypara{Multi-layer Perceptron}
    For multi-layer-perceptron classifier, we use Adam optimizer and Relu activation function. And set the hidden layer size and learning rate to $128$ and $0.001$, respectively.
\end{itemize}

\section{Implementation of SimpleCNN}
\label{app:cnn_structure}

The architecture of our SimpleCNN is illustrated in \autoref{tab:cnn_structure}.
We train the SimpeCNN for $100$ epochs, and use SGD optimizer with learning rate of 0.001.

\begin{table}[!h]
    \centering
    \caption{SimpleCNN structure and hyperparameter. For the MNIST dataset, input\_channel $C_i=1$, image width $W$ and height $H$ are both 28. The kernel\_size of convolution layer $K_c$ and Max-pooling layer $K_m$ are 3 and 2, respectively.}
    \vspace{-0.3cm}
    \resizebox{0.9\linewidth}{!}{
    \setlength{\tabcolsep}{0.3em}
    \renewcommand{\arraystretch}{1.3}
    \begin{tabular}{c c}
    \toprule
    \textbf{Layer} & \textbf{Hyperparameters} \\ 
    \toprule
    Conv2D\_1 & ($C_i$, 32, $K_c$=3, 1) \\
    \rowcolor{mygray}
    Relu & - \\
    Conv2D\_2 & (32, $H$, $K_c$, 1) \\
    \rowcolor{mygray}
    Maxpolling2D & $K_m$=2\\
    Dropout\_1 & (0.25) \\
    \rowcolor{mygray}
    Flatten & 1 \\
    Linear\_1 & ($H\times(W/2-K+1)\times(H/2-K+1)$, 128) \\
    \rowcolor{mygray}
    Relu & - \\
    Dropout\_2 & 0.5\\
    \rowcolor{mygray}
    Linear\_2 & (128, \#classes)\\
    Softmax & dim=1\\
    \bottomrule
    \end{tabular}
    }
    \label{tab:cnn_structure}
\end{table}

\end{document}